\documentclass[fleqn,usenatbib]{mnras}

\usepackage[T1]{fontenc}
\usepackage{ae,aecompl}

\usepackage{graphicx}	
\usepackage{amsmath}	
\usepackage{amssymb}	
\usepackage{xcolor} 
\usepackage{cuted}
\usepackage{xspace}
\usepackage{tabularx}
\usepackage{rotating}
\usepackage{lscape}

\newcommand{\asloth}{\textsc{a-sloth}\xspace}
\newcommand{\ctp}{\textit{Caterpillar}\xspace}
\newcommand{\msun}{{\rm M}_\odot}
\newcommand{\feh}{\mathrm{[Fe/H]}}

\title[Comparing models and observations with ML]{Comparing simulated Milky Way satellite galaxies with observations using unsupervised clustering}
\author[L.-H. Chen et al.]{
Li-Hsin Chen$^{1,2}$, Tilman Hartwig$^{3,4,5}$, Ralf S. Klessen$^{1,6}$, and Simon C. O. Glover$^{1}$
\\
$^{1}$Universit\"{a}t Heidelberg, Zentrum f\"{u}r Astronomie, Institut f\"{u}r Theoretische Astrophysik, Albert-Ueberle-Str.\ 2, \\ 69120 Heidelberg, Germany\\
$^{2}$International Max Planck Research School for Astronomy and Cosmic Physics at the University of Heidelberg (IMPRS-HD), \\ Königstuhl 17, D-69117 Heidelberg, Germany \\
$^{3}${Department of Physics, School of Science, The University of Tokyo, Bunkyo, Tokyo 113-0033, Japan}\\
$^{4}${Institute for Physics of Intelligence, School of Science, The University of Tokyo, Bunkyo, Tokyo 113-0033, Japan}\\
$^{5}$Kavli Institute for the Physics and Mathematics of the Universe (WPI), The University of Tokyo Institutes for Advanced Study, \\ The University of Tokyo, Kashiwa, Chiba, 277-8583, Japan\\
$^{6}$Universit\"{a}t Heidelberg, Interdisziplin\"{a}res Zentrum f\"{u}r Wissenschaftliches Rechnen, Im Neuenheimer Feld 225, \\ D-69120 Heidelberg, Germany
}
\date{}

\begin{document}
\label{firstpage}
\pagerange{\pageref{firstpage}--\pageref{lastpage}}
\maketitle

\begin{abstract}
We develop a new analysis method that allows us to compare multi-dimensional observables to a theoretical model. The method is based on unsupervised clustering algorithms which assign the observational and simulated data to clusters in high dimensionality. From the clustering result, a goodness of fit (the p-value) is determined with the Fisher-Freeman-Halton test. 
We first show that this approach is robust for 2D Gaussian distributions. We then apply the method to the observed MW satellites and simulated satellites from the fiducial model of our semi-analytic code \asloth. We use the following 5 observables of the galaxies in the analysis: stellar mass, virial mass, heliocentric distance, mean stellar metallicity $\feh$, and stellar metallicity dispersion $\sigma_\feh$. 
A low p-value returned from the analysis tells us that our \asloth fiducial model does not reproduce the mean stellar metallicity of the observed MW satellites well. We implement an ad-hoc improvement to the physical model and show that the number of dark matter merger trees which have a p-value > 0.01 increases from 3 to 6. 
This method can be extended to data with higher dimensionality easily. We plan to further improve the physical model in \asloth using this method to study elemental abundances of stars in the observed MW satellites.
\end{abstract}

\begin{keywords}
methods: data analysis -- galaxies: dwarf
\end{keywords}

\section{Introduction}
\label{sec:ml_intro}
There are many dwarf galaxies (with stellar mass~$ < 10^{10}\msun$) surrounding the Milky Way (MW). It is important to understand how these dwarf galaxies form and evolve in order to understand the evolution of the MW system. In the past decades, researchers have devoted considerable effort to the observational study of these dwarf galaxies  \citep{Belokurov:2010aa,Drlica-Wagner:2015aa,Koposov:2015aa,Ji:2021aa}.
Different properties of them have been reported: from the observed populations \citep{Koposov:2009aa,Munoz:2018aa}, the magnitudes, the heliocentric distances, the sizes, and the stellar velocity dispersions \citep[][and the references therein]{McConnachie:2012aa,Simon:2019aa,Wang:2021aa}, to stellar dynamics \citep{Kirby:2017aa,McConnachie:2020aa,Battaglia:2022aa,Battaglia:2022ab}, detailed chemical information \citep{Ji:2016aa,Reichert:2020aa,Yoon:2020aa,Chiti:2018aa,Chiti:2022aa}, and the star formation history \citep{Weisz:2014aa,Gallart:2021aa}.

From a theoretical perspective, numerical simulations and semi-analytic modelling of the formation and evolution of dwarf galaxies have been carried out in the past decades \citep{Ricotti:2005aa,Font:2011aa,Jeon:2017aa,Sanati:2022aa}. Among numerous works, various observables or physical quantities are used to calibrate the models. For example, \citet{Starkenburg:2013aa} reproduced the luminosity function and spatial distributions of the MW satellites. \citet{Salvadori:2015aa} successfully reproduced the metallicity distribution functions and star formation histories of the dwarf galaxies with their semi-analytic model. \citet{Wheeler:2019aa} matched their simulated dwarf galaxies with the observed stellar mass-to-halo mass relation and the 2D half-stellar-mass radii. Other works aim to study individual dwarf galaxies with higher numerical resolution. For instance, \citet{Safarzadeh:2017aa} studied the r-process enrichment in ultra-faint dwarf galaxies and found one of their simulated haloes being similar to Reticulum II. \citet{Romano:2019aa} simulated an isolated dwarf galaxy to understand the importance of stellar feedback in the formation of Bo\"otes I. These works covered different aspects of the formation and evolution of dwarf galaxies. However, a comprehensive understanding is still missing due to the complexity involved in taking all of the different physical processes into account.

The amount of data coming from either observations or simulations increases drastically as observational and computational technology continues to improve. To maximise the information gain from this data, machine learning has been proved to be a powerful tool in many astronomical fields. To give a few examples,  \citet{Garcia-Dias:2018aa} used K-Means unsupervised clustering to classify over 150,000 spectra. \citet{Reis:2019aa} developed a data visualization portal to help researchers spot anomalies in high dimensional astronomical data and dimensionality reduction. \citet{Logan:2020aa} used Hierarchical Density-Based Spatial Clustering of Applications with Noise to distinguish stars, galaxies and quasars. \citet{Ksoll:2021aa,Ksoll:2021ab} used the algorithm \textsc{ransac} to determine the reddening properties of massive stars in giant molecular clouds and built a catalog for $>$ 450,000 stars. \citet{Kang:2022aa} used conditional invertible neural network to study properties of young massive stars with emission lines. \citet{Wang:2022aa} utilized a convolutional neural network to recover the cosmic microwave background signal.

In this work, we introduce a different analysis method to help us analyse the results produced by our semi-analytic code \asloth 
\citep{Hartwig:2022aa,Magg:2022ac}. 
\asloth has been used to make predictions of Population~III survivors in the MW, \citep{Hartwig:2015aa}, Population~III supernovae rate \citep{Magg:2016aa}, to study the inhomogeneous mixing \citep{Tarumi:2020ab}, and the properties of the MW satellites \citep{lhc:2022aa}. 
In these earlier works, the calibration of the physical model or the consistency check was carried out by comparing individual properties of simulated galaxies and observed galaxies. However, we noticed that it was difficult to find the best parameter combination by doing so. For instance, if we tune the parameter to move the simulated MW stellar mass closer to the observed value, we actually observe a larger difference between the simulated MW cold gas mass and the observed value \citep[see Fig.~2 in][for illustration]{lhc:2022aa}. To reduce the effort of cross checking the consistency between simulated properties and observed ones, a new analysis method is developed to compare multiple properties of the observed and simulated galaxies simultaneously with the help of unsupervised clustering algorithms.
\section{Method}
\label{sec:ml_met}
We aim to compare our simulated data with observables in high-dimensional data space. One value should determine whether the simulated data and the observational data come from the same underlying distribution. In 1D, a renowned test to compare two data distributions is the Kolmogorov–Smirnov test \citet[K-S]{Kolmogorov:1933aa} test. It computes the maximal Euclidean distances between the two distributions and return a p-value, which helps us reject the null hypothesis. Generalisation of the K-S test to 2D data is shown in the 1980s \citep{Peacock:1983aa,Fasano:1987aa}. 
In the following sections, we first describe the observed galaxies and the simulated galaxies used in the analysis. We then describe how we determine whether the physical model in \asloth is successful at reproducing the observed properties of the MW satellites.

Among numerous properties of the MW satellites, we are most interested in the stellar mass and chemical composition of the MW satellites. We choose to analyse the following five quantities in this work: stellar mass ($M_*$), virial mass ($M_\mathrm{vir}$), distance to the Sun ($D_\odot$), the mean stellar metallicity ($\feh$) and the stellar metallicity dispersion ($\sigma_\mathrm{\feh}$).
From the simulation, we do not have information on the spatial distribution of baryons within each halo. Therefore, we leave properties that require further assumptions regarding this, such as the half-light radius, velocity dispersion, or radial velocity for future work.

\subsection{Observational data}
\label{sec:ml_obs}
We collect the following values of the observed MW satellites from \citet{Simon:2019aa} and the references therein: the V-band absolute magnitude $M_\mathrm{V}$, distance to the Sun $D_\odot$, the stellar velocity dispersion $\sigma_{*}$, and the half light radius $R_\mathrm{half}$. 
We compute $\langle \mathrm{[Fe/H]} \rangle$ and $\sigma_\mathrm{[Fe/H]}$ from individual detections in the SAGA database\footnote{\href{http://sagadatabase.jp/}{http://sagadatabase.jp/}} \citep{Suda:2008aa,Suda:2017aa}, except for Horologium I, Tucana III, Grus I, and Pisces II. These four galaxies have $\leq 3$ detections available in the SAGA database. Therefore, we obtain $\langle \mathrm{[Fe/H]} \rangle$ and $\sigma_\mathrm{[Fe/H]}$ for these galaxies again from \citet{Simon:2019aa}. For Reticulum II, we add 5 newly detected stars reported by \citet{Chiti:2022aa} and compute the mean [Fe/H] and the standard deviation along with the data from the SAGA database. To obtain the stellar mass of the observed satellites from $M_\mathrm{V}$, we simply assume a stellar mass-to-light ratio of 1 in units of ($\msun/L_\odot$) \citep{McConnachie:2012aa}.



It is observationally challenging to estimate the virial masses of the dark matter haloes in which the observed galaxies reside, because we cannot observe dark matter directly. There is also no clear boundary of the dark matter halo. 
Some researchers utilise the observed stellar velocity dispersion and model the dark matter haloes of observed MW satellites \citep{Munoz:2006aa,Walker:2007aa,Chiti:2021ac}. 
\citet{Errani:2018aa} provided an estimate of mass enclosed in $1.8 R_\mathrm{half}$ for dwarf spheroidal galaxies 
\begin{equation}
\label{eq:mhaloerrani}
    M(<1.8 R_\mathrm{half}) = 3.5 \times 1.8 R_\mathrm{half} \sigma^2_{*} \mathrm{G}^{-1},
\end{equation}
where G is gravitational constant and $R_\mathrm{half}$ is the half-light radius. 
We adopt virial masses of the observed MW satellites if they are provided in the literature, otherwise we simply take $M_\mathrm{vir} = 10M(<1.8 R_\mathrm{half})$. This factor of 10 is relatively arbitrary. From the 8 galaxies that have literature values, the difference between $M_\mathrm{vir}$ and $M(<1.8 R_\mathrm{half})$ is on the order of 10. Therefore we adopt 10 as the fiducial value but this is to be improved with more precise computation of the virial mass for each observed MW satellite.
The physical quantities of observed MW satellites are listed in Table~\ref{tab:ml_obs}. 
\begin{landscape}
\begin{table}
\centering
\begin{tabularx}{0.8\paperheight}{ *{1}{p{2.5cm}}|*{6}{p{1.3cm}}*{1}{p{2.9cm}}*{1}{p{1.3cm}}|*{1}{p{2.cm}} } 
    \hline
    (1) & (2) & (3) & (4) & (5) & (6) & (7) & (8) & (9) & (10) \\
    
    Galaxy & $M_\mathrm{*}$ & $D_\odot$ & $\sigma_{*}$ & $\langle \mathrm{[Fe/H]} \rangle$ & 
    $\sigma_\mathrm{[Fe/H]}$ & $R_\mathrm{half}$ & $M_\mathrm{halo} (<1.8 R_\mathrm{half})$ & $M_\mathrm{vir}$ & References \\
    
    & log$_{10}\msun$ & kpc & km s$^{-1}$ & & & pc & log$_{10}\msun$ & log$_{10}\msun$ & \\
    \hline
    Bootes I & 4.33 & 66.9 & 4.6 & -2.60 & 0.42 & 191 & 6.77 & 7.00 & 1,1,1,2,2,1,,5  \\ 
    Bootes II & 3.10 & 42.0 & 10.5 & -2.34 & 0.65 & 42 & 6.83 & 7.83 & 1,1,1,2,2,1,, \\
    Canes Venatic I & 5.41 & 211.0 & 7.6 & -1.91 & 0.54 & 211 & 7.25 & 8.25 & 1,1,1,2,2,1,, \\
    Canes Venatic II & 4.00 & 160.0 & 4.6 & -2.21 & 0.60 & 162 & 6.70 & 7.70 & 1,1,1,2,2,1,, \\
    Carina & 5.70 & 106.0 & 6.6 & -1.46 & 0.55 & 311 & 7.30 & 8.30 & 1,1,1,2,2,1,,4 \\
    Coma Berenices & 3.63 & 42.0 & 4.6 & -2.72 & 0.36 & 69 & 6.33 & 7.33 & 1,1,1,2,2,1,, \\
    Draco & 5.47 & 82.0 & 9.1 & -1.97 & 0.46 & 231 & 7.44 & 9.60 & 1,1,1,2,2,1,,4 \\
    Fornax & 7.26 & 139.0 & 11.7 & -1.10 & 0.49 & 792 & 8.20 & 9.00 & 1,1,1,2,2,1,,4 \\
    Grus I & 3.31 & 120.0 & 2.9 & -1.42 & 0.41 & 28 & 5.53 & 6.53 & 1,1,1,1,1,1,, \\
    Hercules & 4.25 & 132.0 & 5.1 & -2.43 & 0.40 & 216 & 6.92 & 7.92 & 1,1,1,2,2,1,, \\
    Horologium I & 3.42 & 87.0 & 4.9 & -2.76 & 0.17 & 40 & 6.15 & 7.15 & 1,1,1,1,1,1,, \\
    Leo I & 6.63 & 254.0 & 9.2 & -1.32 & 0.34 & 270 & 7.52 & 9.00 & 1,1,1,2,2,1,,4 \\
    Leo II & 5.82 & 233.0 & 7.4 & -1.56 & 0.40 & 171 & 7.14 & 8.60 & 1,1,1,2,2,1,,4 \\
    Leo IV & 3.92 & 154.0 & 3.3 & -2.47 & 0.50 & 114 & 6.26 & 7.26 & 1,1,1,2,2,1,, \\
    Pisces II & 3.61 & 183.0 & 5.4 & -2.45 & 0.48 & 60 & 6.41 & 7.41 & 1,1,1,1,1,1,, \\
    Reticulum II & 3.51 & 31.6 & 3.3 & -2.88 & 0.52 & 51 & 5.91 & 6.91 & 1,1,1,2+3,2+3,1,, \\
    Sagittarius & 7.32 & 26.7 & 9.6 & -0.54 & 0.31 & 2662 & 8.56 & 9.56 & 1,1,1,2,2,1,, \\
    Sculptor & 6.25 & 86.0 & 9.2 & -1.86 & 0.61 & 279 & 7.54 & 9.00 & 1,1,1,2,2,1,,4 \\
    Segue 1 & 2.44 & 23.0 & 3.7 & -2.52 & 0.88 & 24 & 5.68 & 6.68 & 1,1,1,2,2,1,, \\
    Segue 2 & 2.71 & 37.0 & 2.2 & -2.24 & 0.40 & 40 & 5.45 & 6.45 & 1,1,1,2,2,1,, \\
    Sextans & 5.50 & 95.0 & 7.9 & -2.12 & 0.54 & 456 & 7.62 & 8.48 & 1,1,1,2,2,1,,4 \\
    Triangulum II & 2.56 & 28.4 & 3.4 & -2.43 & 0.49 & 16 & 5.43 & 6.43 & 1,1,1,2,2,1,, \\
    Tucana II & 3.48 & 58.0 & 8.6 & -2.94 & 0.29 & 121 & 7.12 & 8.12 & 1,1,1,2,2,1,, \\
    Tucana III & 2.52 & 25.0 & 1.2 & -2.42 & 0.19 & 37 & 4.89 & 5.89 & 1,1,1,1,1,1,, \\
    Ursa Major & 3.97 & 97.3 & 7.0 & -2.04 & 0.56 & 295 & 7.32 & 8.32 & 1,1,1,2,2,1,, \\
    Ursa Major II & 3.70 & 34.7 & 5.6 & -2.13 & 0.68 & 139 & 6.80 & 7.80 & 1,1,1,2,2,1,, \\
    Ursa Minor & 5.53 & 76.0 & 9.5 & -2.04 & 0.48 & 405 & 7.73 & 8.73 & 1,1,1,2,2,1,, \\
    William I & 3.08 & 45.0 & 4.0 & -1.40 & 0.40 & 33 & 5.89 & 6.89 & 1,1,1,2,2,1,, \\ 
    \hline
\end{tabularx}
\caption[Physical quantities of the observed MW satellites used in the analysis.]{Physical quantities of the observed MW satellites that are used in the analysis. From left to right: Galaxy name, stellar mass, heliocentric distance, velocity dispersion, mean stellar metallicity, scatter of stellar metallicity, half light radius, halo mass with in 1.8 $R_\mathrm{half}$, virial mass estimate. Refereces: (1) \citet{Simon:2019aa} and the references therein, (2) {SAGA database} \citep{Suda:2008aa}, (3) \citet{Chiti:2022aa}, (4) \citet{Walker:2007aa}, (5) \citet{Munoz:2006aa}. Note that the stellar mass is derived from V-band magnitude assuming stellar-to-light ratio of 1 and the halo mass with in 1.8 $R_\mathrm{half}$ is derived from Eq.~\ref{eq:mhaloerrani}. If there is no reference for the virial mass, we simply adopt $M_\mathrm{vir} = 10M_\mathrm{halo} (<1.8 R_\mathrm{half})$.}
\label{tab:ml_obs}
\end{table}
\end{landscape}

Note that the Small Magenllanic Cloud (SMC) and the Large Magellanic Cloud (LMC) are excluded in this analysis because there is no implementation of Type Ia Supernovae in \asloth, which is required to explain the chemical features of SMC and LMC \citep{Tsujimoto:1995aa,Rolleston:2003aa,VanderSwaelmen:2013aa}.
In addition, we do not consider Leo T in the sample because it is located at $> 400\,$kpc from the MW and the merger trees that we use only consider galaxies within the virial radius ($\sim 300\,$kpc) of the MW as satellites (Sec.~\ref{sec:ml_simsat}). 

Finally, we apply a selection function to the galaxies based on their heliocentric distances and V-band absolute magnitudes \citep{Koposov:2009aa},  
\begin{equation}
    \mathrm{log}_{10}(D_\odot/1 \mathrm{kpc}) < 1.1 - 0.228 M_\mathrm{V},
\end{equation}
to account for the observational incompleteness and to make fair comparison with the simulation.

\subsection{Simulated MW satellites}
\label{sec:ml_simsat}
We generate simulated MW satellites by running the fiducial model of \asloth \citep{Hartwig:2022aa,Magg:2022ac}. We briefly summarise the model here.
\asloth is a semi-analytic model that takes dark matter merger trees as input. It assigns the baryonic content inside the haloes based on the included physical models. The physical processes include stochastic star formation of metal-free and metal-poor stars, kinematic and chemical feedback from Type II SNe and Pair instability SNe, tracing of elemental abundances of the SNe yields in the ISM and individual stars. We utilise 30 dark matter merger trees from the \ctp project \citep{Griffen:2016aa}. They selected MW-like haloes based on the following criteria: 
\begin{enumerate}
  \item \noindent Virial mass is in the range of $0.7 \times 10^{12}\msun$ $\leq M_\mathrm{vir} \leq 3 \times 10^{12}\msun$. 
  \item \noindent There is no halo with $M_\mathrm{vir} \geq 7 \times 10^{13}\msun$ within 7~Mpc. 
  \item \noindent There are no other haloes with $M_\mathrm{vir} \geq 0.5 \times M_{\rm main}$ within 2.8 Mpc, where $M_{\rm main}$ is the virial mass of the main halo. 
\end{enumerate}
Note that only galaxies that are within $300\,$kpc from the MW are considered as satellites. 
We apply the same selection function as in Sec.~\ref{sec:ml_obs} to filter out small, distant galaxies. 

Since the location of the Sun is not known from the dark matter only simulation, we randomly pick the solar position in the MW at a radius of 8.5 kpc \citep{Koposov:2009aa} and compute the distance to the Sun for \asloth simulated satellites with 
\begin{equation}
    D_\odot = \sqrt{8.5^{2} + D_\mathrm{MW}^2 - 2 \times 8.5 \times D_\mathrm{MW} \times \mathrm{cos(\phi)}},
\end{equation}
where $D_\mathrm{MW}$ is the distance to the MW center (in kpc) from the simulations,  $\mathrm{cos(\phi)}$ is a random number uniformly distributed between -1 and 1, and $\phi$ is the angle between the vectors from the MW center to Sun and to the satellite. 
We only consider satellites with stellar masses $< 10^8\msun$ since we do not aim to compare with SMC and LMC.
Due to the uncertainty in the solar position, we determine the final p-value of our fiducial model by running the same analysis 100 times and take the geometric mean of these 100 p-values as the final p-value.

\subsection{Unsupervised clustering algorithm}
\label{sec:ml_clust}
The fiducial unsupervised clustering that we use is the Agglomerative clustering. 
Agglomerative is a bottom-up hierarchical clustering algorithm. It starts by pairing the data points and then merge the pairs into clusters, eventually leading to a tree-like diagram, the dendrogram \citep{scikit-learn}. In principle, the algorithm does not aim to find ``n clusters", therefore, a pre-assigned number of clusters is not required. When the user assigns the number of clusters they want to find, the algorithm stops the merging when the number of clusters is reached. 
Data points are then returned with labels, indicating which cluster they belong to. We discuss other unsupervised clustering algorithms and the dependence of the result on the number of clusters in Sec.~\ref{sec:ml_results}.

\subsection{Goodness of fit}
\label{sec:ml_pvalue} 
Once the clusters are found by the unsupervised clustering algorithm, we construct a contingency table that shows how many observed galaxies and simulated galaxies are assigned to the clusters. 
To determine the p-value from the contingency table, there is the Pearson's chi-squared test \citep{Peasron:1916aa}. It computes the differences between the expected values and the actual outcome. The difference then corresponds to a p-value. There is no limitation on data dimensionality to apply the Pearson's chi-squared test. However, one has to make sure that the expected frequency is larger than 5. It is therefore not suitable to use the Pearson's chi-squared test when the number of data points is small.
Since we aim to compare two datasets in high dimensional space and there are only a handful of observed MW satellites, it is likely that there are very few observed MW satellites in the clusters, leading to small expected frequency. 
Therefore, we decide to use the Fisher-Freeman-Halton exact test \citep{Fisher:1934aa,Freeman:1951aa}, as the fiducial test. It computes the probability of the observed outcome based on the ratio of the sizes of the two datasets. 
The Fisher-Freeman-Halton exact test is not dependent on the number of data points in each cluster and has no limitation of data dimensionality. 
If the two datasets do not come from the same underlying distributions (whether it's Gaussian-like distribution or the real data), we expect the unsupervised clustering algorithm to assign data points from different subsets to different clusters and a low p-value from the Fisher-Freeman-Halton exact test. This is discussed in more details with an example in Sec.~\ref{sec:ml_gau}.

\section{Results}
\label{sec:ml_results}
In this section, we first show results from test cases where we use datasets sampled from Gaussian distributions to show that our method works. We compare results from different unsupervised clustering algorithms and discuss the dependence of our results on the number of clusters and justify our choice of fiducial value. 
Next, we present the main results from our analysis: the p-value of our fiducial model where we consider all MW satellites in 30 \ctp trees as one dataset (the Ensemble) and the p-values where we consider MW satellites in individual \ctp trees as individual datasets.

\subsection{Application to Gaussian distributions}
\label{sec:ml_gau}
To illustrate that our method can distinguish good models from bad models, we test it with two-dimensional Gaussian distributions first. 
Most importantly, we are interested in the dependence of the p-value on the number of clusters. 
In Fig.~\ref{fig:ncdep_gau} we show p-value vs.\ the number of clusters for three test cases. In case 1, we sample 2 subsets from two identical two-dimensional Gaussian distributions. 
In case 2 (3), we shift one of the Gaussian distributions by 0.5 (1.0) $\sigma$ before we sample data points from it. 
Subset 1 has 30 data points and subset 2 has 5570 data points, which is roughly the ratio of observed satellites to simulated satellites that we will use later.

\begin{enumerate}
    \item \textbf{Different unsupervised algorithms} \\
    Here we compare different unsupervised clustering algorithms: KMeans, Agglomerative hierarchical clustering, Spectral clustering, and Balanced Iterative Reducing and Clustering using Hierarchies (BIRCH). 

    We briefly summarise the concept in these algorithms: in KMeans, the user assigns the desired number of clusters. The algorithm assigns the centres of the clusters and iterates to minimise the variance within each cluster. In Agglomerative hierarchical clustering, each data point starts as an individual clusters and clusters are then merged based on the distance between them. Spectral clustering first computes the similarity matrix, which estimates the similarity between the data points from the original input data. It then clusters the data points with higher similarities using existing methods such as KMeans. 
    BIRCH does not use the distances in the original parameter space but first builds clustering features (CFs) for the input data. These CFs are then organised into a height-balanced CF tree. It then applies the Agglomerative algorithm to cluster the leaves in the CF trees. The description of these unsupervised clustering algorithms and their usage can be found in \citet{scikit-learn}. 

    From Table~\ref{tab:fourmethod}, we observe that KMeans finds clusters with even sizes, whereas BIRCH finds clusters with the most uneven sizes. The p-values obtained from these four clustering algorithms range from $10^{-5}$ to $10^{-9}$. Although the range in values is large, all of the p-values are sufficiently small such that we can reject the null hypothesis that the data come from the same underlying distribution. 
    Although KMeans is the easiest-to-understand algorithm, it has some limitations. KMeans is not good at handling outliers or identifying clusters with non-convex shapes and there is an assumption of the number of clusters to be found. Therefore, as mentioned in Sec.~\ref{sec:ml_clust}, we choose Agglomerative as the fiducial clustering algorithm and apply it to our data.

\begin{table}
    \centering
    \begin{tabular}{l|l|l|l|l|l}
        \textbf{KMeans}        & Cl. 1 & Cl. 2 & Cl. 3 & Cl. 4 & \textbf{p-value} \\
        \hline
        Small subset & 22    & 0     & 7      & 1    & 1.33$\times 10^{-7}$ \\
        Large subset  & 1668  & 1606  & 1398   & 1298 & \\
        \hline
        \hline
        \textbf{Agglomerative}  &  &  &  &  & \textbf{p-value} \\
        \hline
        Small subset & 8     & 22    & 0     & 0 & 3.85$\times 10^{-8}$ \\
        Large subset & 2203  & 1469  & 1210  & 1088 & \\
        \hline
        \hline
        \textbf{Spectral}  &  &  &  &  & \textbf{p-value} \\
        \hline
        Small subset & 23    & 0     & 6     & 1   & 1.04$\times 10^{-5}$ \\
        Large subset  & 2311  & 1643  & 1050  & 966 & \\  
        \hline
        \hline
        \textbf{BIRCH}  &  &  &  &  & \textbf{p-value} \\
        \hline
        Small subset & 29    & 0     & 0    & 0   & 5.93 $\times 10^{-9}$ \\
        Large subset  & 2572  & 2526  & 712  & 160 & \\          
    \end{tabular}
    \caption[Example of contingency tables]{Exemplary contingency tables: we draw two subsets from two 2D Gaussian distributions and apply four different unsupervised clustering algorithms to find four clusters. The underlying Gaussian distributions are separated by 1\,$\sigma$.
    The listed p-values are computed with the
    null hypothesis that both subsets are drawn
    from the same distribution. All of the unsupervised clustering algorithms yield
    small p-values and allow us to (correctly) reject the null hypothesis.}
    \label{tab:fourmethod}
\end{table}

    \item \textbf{Dependence of p-value on the number of clusters} \\
    The four unsupervised algorithms all allow or require a user-defined number of clusters. Here we show the dependence of the p-value on the number of clusters. When we draw the subsets from two identical Gaussian distributions (case 1), the p-values are similar regardless of the number of clusters. When the Gaussian distributions are separated by 0.5 $\sigma$, we observe a small decrease in the p-value when the number of cluster increases from 2 to 3, but the value stays almost constant afterwards. When the Gaussian distributions are separated by 1.0 $\sigma$, the decrease in p-value continues until 5 clusters and we start to observe distinctive p-values from the 4 unsupervised cluster algorithms. 
    At $1 \,\sigma$ apart, our method returns p-values below $0.01$ which gives us confidence to reject the null hypothesis that the two subsets come from the same underlying distribution. Ideally, the p-value should be independent of the number of clusters. However, if the number of clusters is small, it is likely that most of the data points are in any case assigned to only one or two of the clusters, which could lead to a p-value that is biased towards the higher value. Thus, we choose 5 clusters as the fiducial value.  

\begin{figure}
    \centering
    \includegraphics[width=\columnwidth]{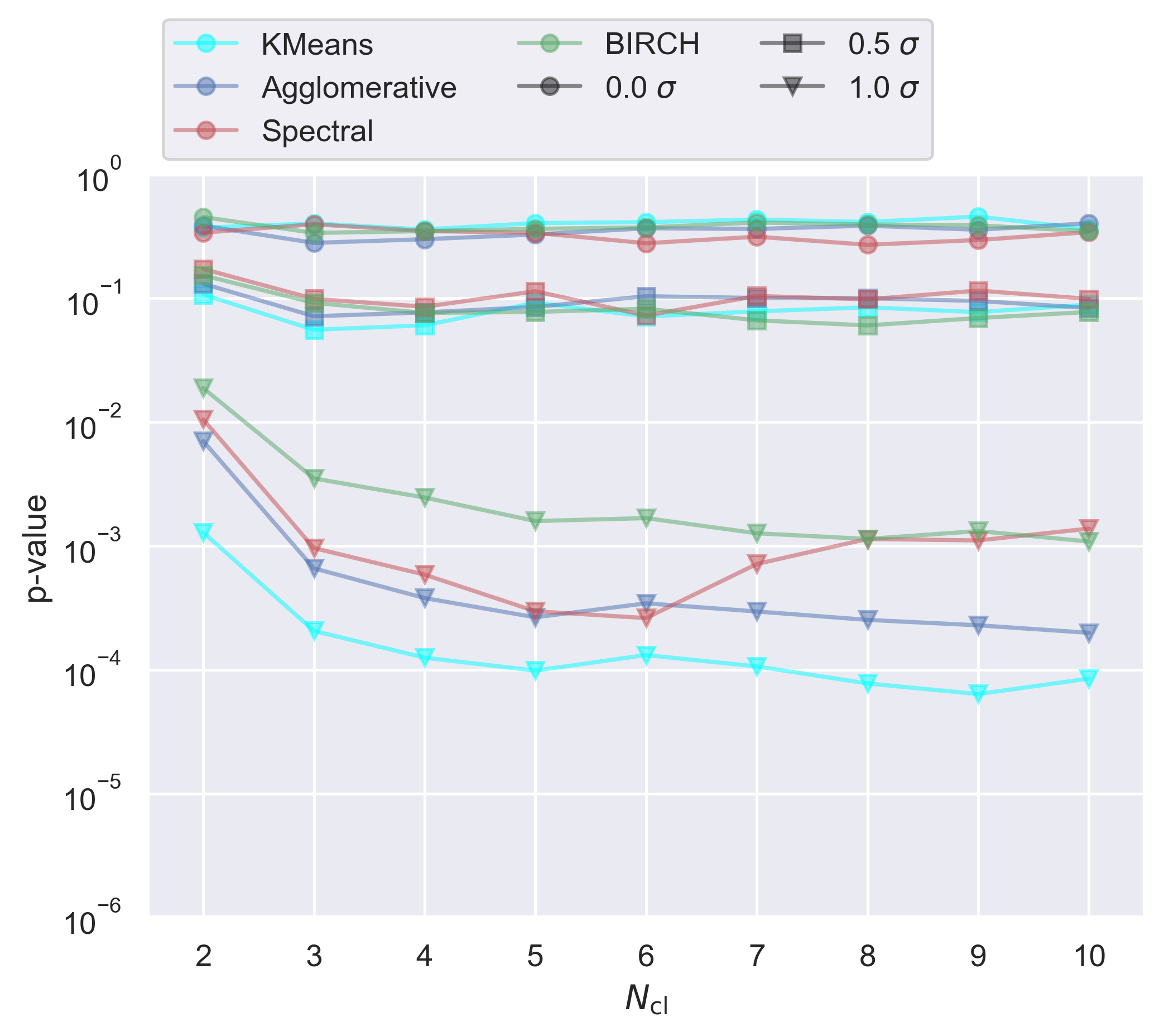}
    \caption[p-value vs.\ number of clusters with different unsupervised clustering methods]{p-value vs.\ number of clusters with different unsupervised clustering algorithms. There are three test cases: In case 1, we draw two subsets from two identical Gaussian distributions. In case 2 (3), we shift one of the Gaussian distributions by 0.5 (1.0) $\sigma$ before sampling the subsets. Subset 1 has 30 data points and subset 2 has 5570 data points, which is roughly the ratio of observed MW satellites to simulated satellites.
    The colours indicate different unsupervised clustering algorithms. The circles, squares, and triangles show results from case 1, 2 and 3 (0, 0.5, and 1.0 $\sigma$), respectively.}
    \label{fig:ncdep_gau}
\end{figure}
\end{enumerate}

\subsection{Clustering result and the p-value from our fiducial model}
\label{sec:ml_fidres}
\begin{figure}
    \centering
    \includegraphics[width=\columnwidth]{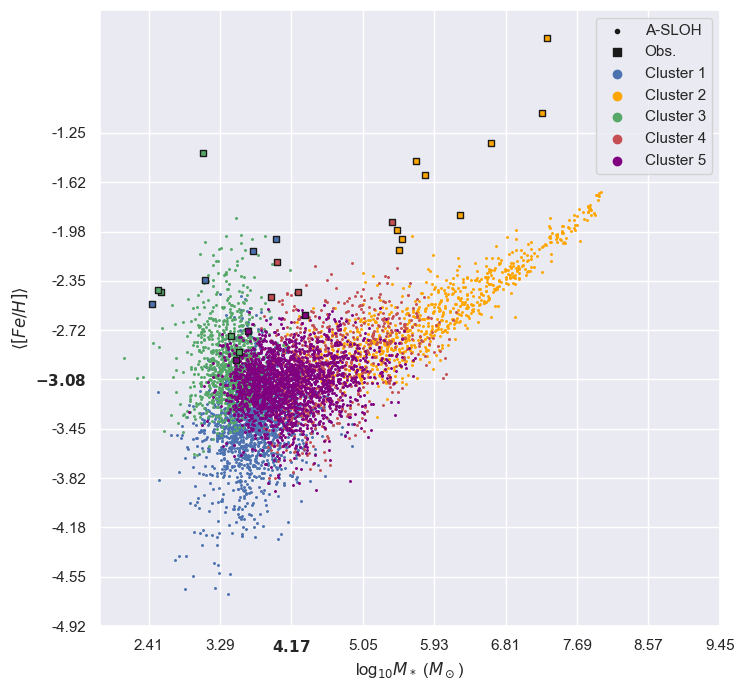}
    \caption[Example of clustering results from \asloth fiducial model]{ An example of unsupervised clustering from the fiducial model of \asloth and the observed satellites, which is projected onto the $\langle\feh\rangle$-${M_*}$ space. The data is normalised in each dimension before we apply the unsupervised clustering and the means are shown in bold font. In this analysis we use Agglomerative with 5 clusters. Galaxies that are classified in different clusters are plotted with different colours. Observed MW satellites are shown in squares and \asloth simulated MW satellites are shown in circles. The p-value for this example is $10^{-2.2}$.}
    \label{fig:fid_clus}
\end{figure}

We show an example of the clustering result from our fiducial model in Fig.~\ref{fig:fid_clus}. All simulated galaxies from 30 \ctp trees are considered (the Ensemble). 
The data in each dimension is normalised before applying the Agglomerative clustering with 5 clusters. 
The clustering result is projected onto the $\langle\feh\rangle$-${M_*}$ space and the mean values are shown in bold font. Galaxies that are assigned to different clusters are shown with different colours. The observed and the \asloth simulated satellites are shown with squares and circles, respectively. 

\begin{figure}
    \centering
    \includegraphics[width=\columnwidth]{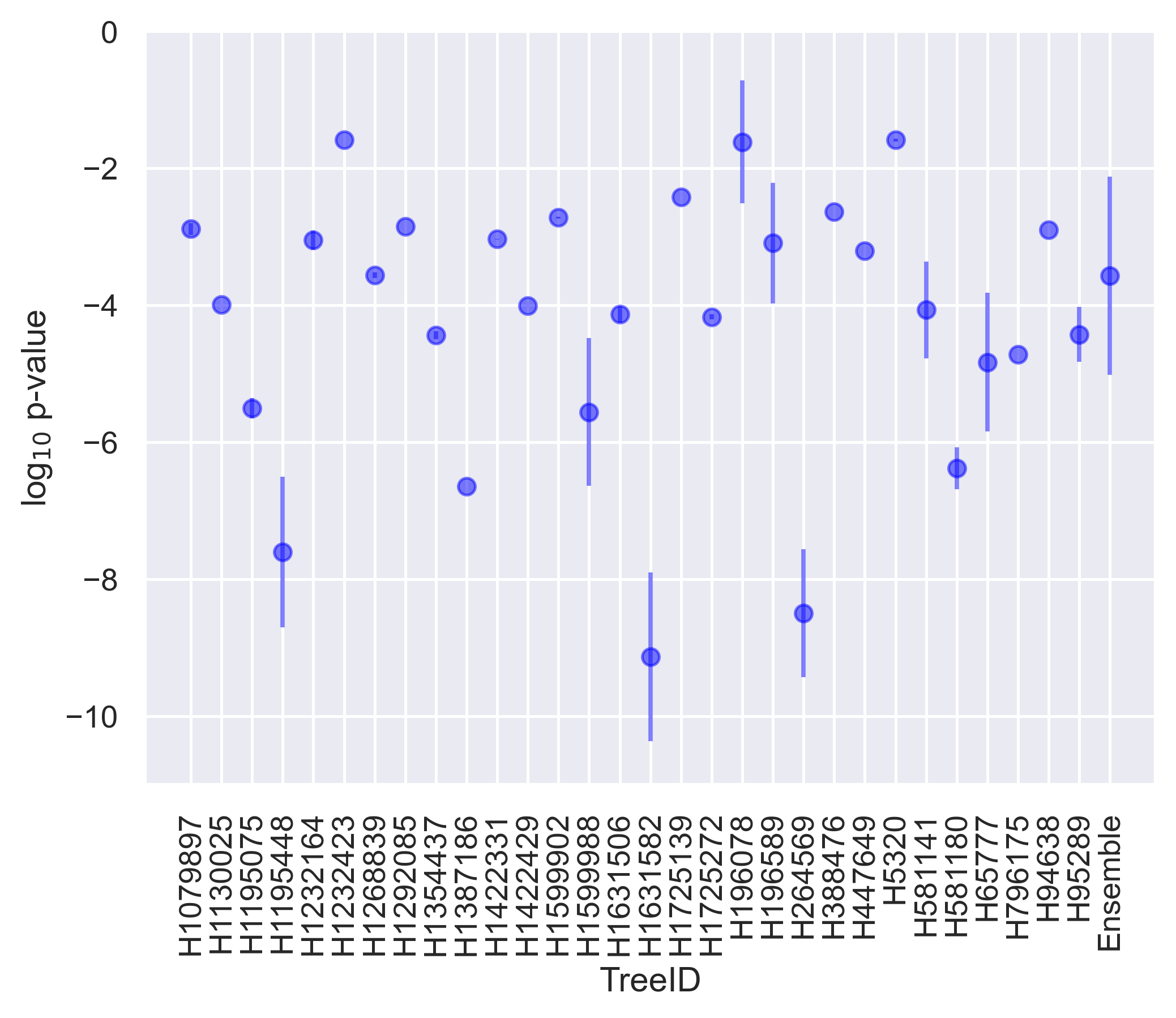}
    \caption[The p-values from 30 \ctp trees]{The p-values from 30 Caterpillar trees as individual datasets and the p-value from the Ensemble (data from 30 trees combined before conducting the analysis). Mean and 1 standard deviation from 100 runs of the analysis are shown with error bars.}
    \label{fig:indi_pv}
\end{figure}

\begin{figure}
    \centering
    \includegraphics[width=\columnwidth]{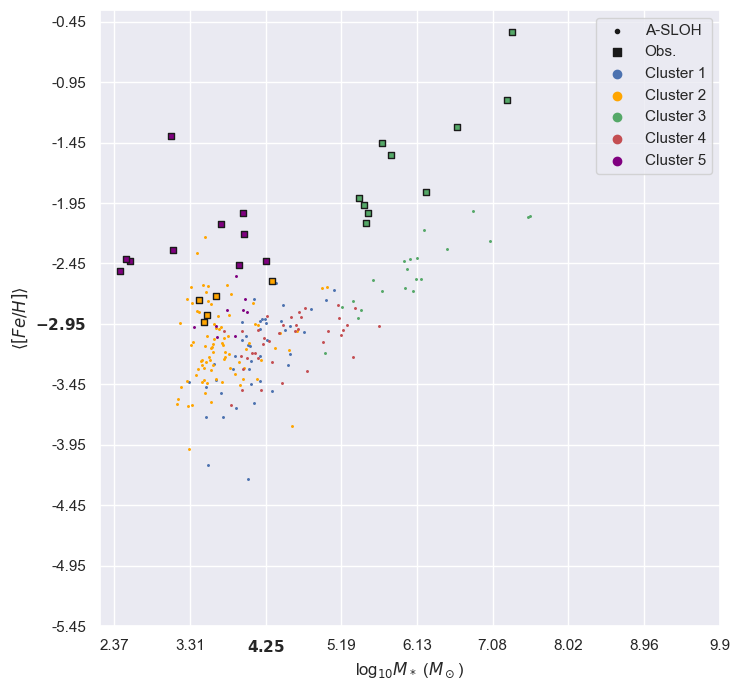}
    \caption[Clustering result projected onto the $\feh$-$M_*$ space for one \ctp tree.]{Similar to Fig.~\ref{fig:fid_clus} but for one specific tree H1631582, which has the lowest p-value among the 30 \ctp trees. Two clusters only consist of simulated satellites and the p-value for this specific run is $10^{-9.6}$.}
    \label{fig:indi_1631582}
\end{figure}

As mentioned in Sec.~\ref{sec:ml_simsat}, we run the same analysis 100 times to take into account of the randomness in the solar position and obtain 100 p-values. 
We compute the geometric mean of the 100 p-values that we obtain with the Ensemble and take it as the final p-value for our fiducial model, which is $10^{-3.5 \pm 1.4}$. 
In Fig.~\ref{fig:indi_pv}, we show the mean p-values with 1 standard deviation of 30 \ctp trees individually along with the p-value from the Ensemble. 
Even though the p-value from the Ensemble is low, we find that the p-values from individual \ctp trees span a wide range. This indicates that some \ctp trees are more similar to the MW than the others. 
For example, in one specific run for Tree H1631582, the algorithm finds two clusters that only consist of the simulated galaxies (Fig.~\ref{fig:indi_1631582}), which leads to a p-value of $10^{-9.6}$.
Further analysis of the merger histories of the individual \ctp trees with high p-values could potentially tell us more about the merger history of the MW. 


\section{Discussion}
\label{sec:ml_dis}
\subsection{Properties of the simulated MW satellites}
\label{sec:ml_fidimp}
\begin{figure}
    \centering
    \includegraphics[width=0.22\paperheight]{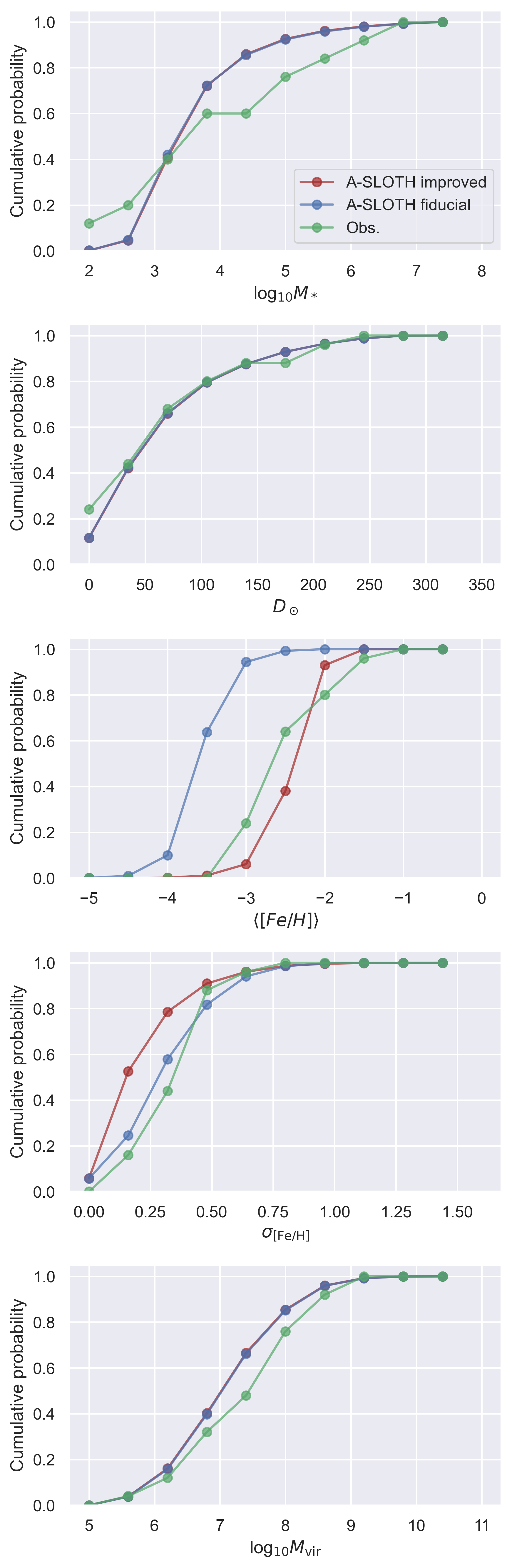}
    \caption[Normalised, cumulative histogram of the five physical quantities used in the analysis]{Normalised, cumulative histogram of the five physical quantities used in the analysis from our fiducial model, improved model, and the observation. From top to bottom: stellar mass, helicocentric distance, mean stellar $\feh$, standard deviation of stellar $\feh$, and the virial mass of the halo. The observation is plotted in green, the \asloth fiducial model is plotted in blue, and the improved model is plotted in brown. }
    \label{fig:hist_fid}
\end{figure}

In this section we discuss the reason why our fiducial model does not reproduce the observables of interest in more details. 
In Fig.~\ref{fig:hist_fid} we show histograms of the five physical quantities that are used in our analysis from both the observed and \asloth simulated satellites (fiducial and improved model). 
The biggest difference between the observed MW satellites and simulated galaxies from the fiducial model lies in the mean stellar [Fe/H], where we observe a difference of $\sim$ 1 dex. 
The overall distribution of the standard deviation of stellar [Fe/H] among the satellites is similar between the observed one and the fiducial model. There are a small number of simulated satellites that have scatter larger than 1 dex among the stars. 

Here we define $N_\mathrm{pass}$ as the number of \ctp trees that have p-values > 0.01. We use it as an indicator of whether \asloth is able to reproduce the observables. In addition to the visual inspection of Fig.~\ref{fig:hist_fid}, we re-run the analysis and exclude one physical quantity at a time to examine which one is most responsible for the inconsistency. The resulting $N_\mathrm{pass}$ is shown in Table~\ref{tab:excl_npass}. It is clear from this table that the mean stellar $\feh$ is crucial. From the \asloth fiducial model, $N_\mathrm{pass} = 24$ if we exclude the mean stellar $\feh$, whereas $N_\mathrm{pass} = 0$ when we exclude any of the other four quantities.

\begin{table}
    \centering
    \begin{tabular}{l|l|l}
        Excluded quantity & $N_\mathrm{pass, fid}$ & $N_\mathrm{pass, imp}$ \\
        \hline
        $M_*$ & 0 & 10 \\
        $D_\odot$ & 0 & 4 \\
        $\langle \feh \rangle$ & 24 & 24 \\
        $\sigma_{\feh}$ & 0 & 3 \\
        $M_\mathrm{vir}$ & 0 & 2 \\
        \hline
        All 5 quantities considered & 3 & 6 \\
    \end{tabular}
    \caption{We run the same analysis but exclude one physical property at a time and show the number of \ctp trees that have p-values > 0.01 for the fiducial model and the improved model ($N_\mathrm{pass, fid}$ and $N_\mathrm{pass, imp}$). For comparison, we show $N_\mathrm{pass}$ from the fiducial and improved models if we consider all five physical quantities in the analysis.}
    \label{tab:excl_npass}
\end{table}

In the fiducial model of \asloth, we assume that the metals ejected from the halo mix with the inter-galactic medium (IGM) homogeneously and instantaneously, where we compute a mixing volume that is monotonically increasing over time. In reality, the metals expand outwards from the halo to the IGM. The newly injected metals should therefore stay in proximity to the halo, leading to a gradient in the radial metallicity profile of the IGM. Without exact spatial information of the gas, we are not able to model this radial profile properly.
We here implement an ad-hoc solution where we assume that the re-accreted gas has a metallicity 8 times higher than the average IGM value.

\begin{figure}
    \centering
    \includegraphics[width=\columnwidth]{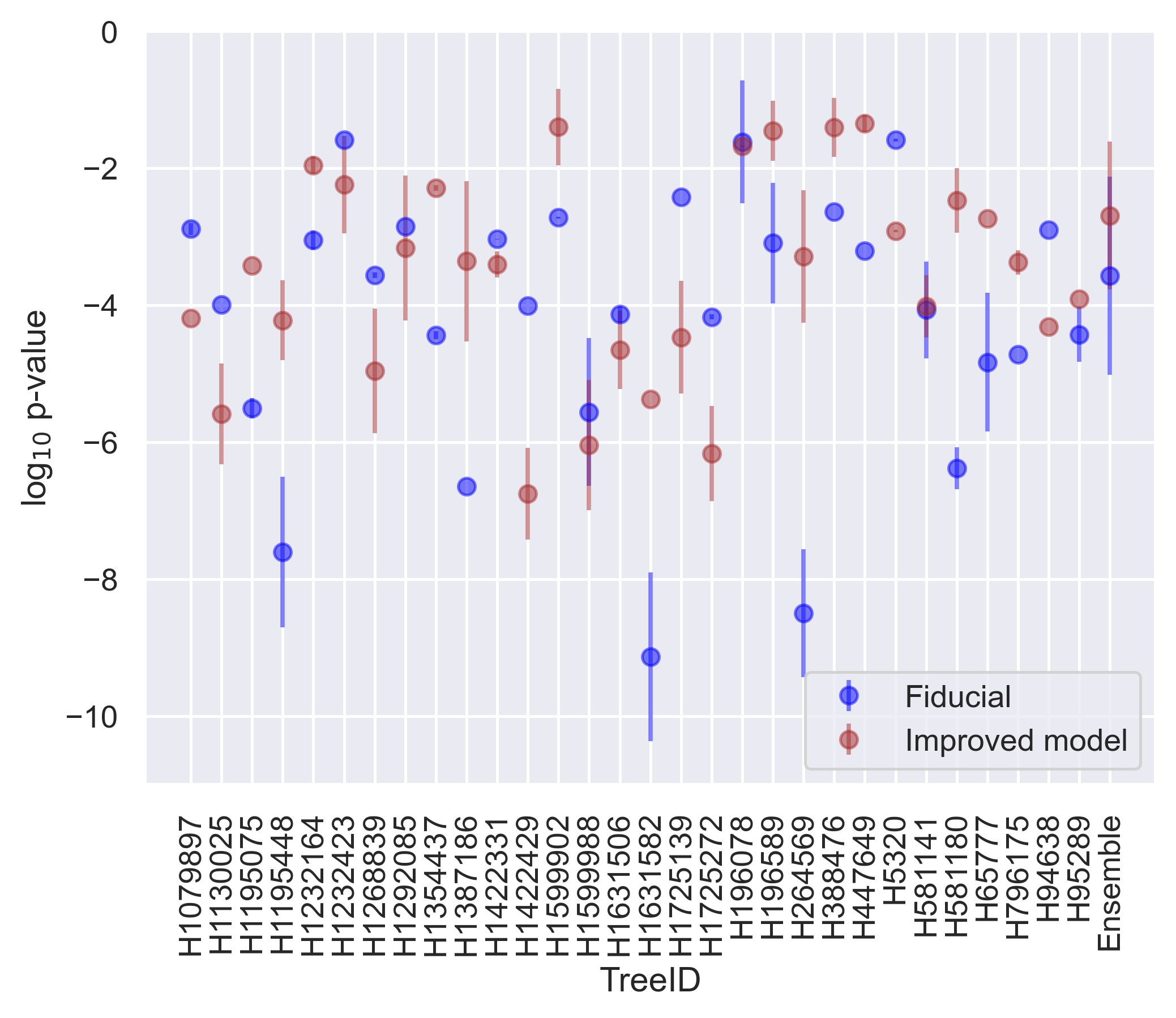}
    \caption{Similar to Fig.~\ref{fig:indi_pv} but we add the p-values from the improved model in brown. The p-values from the fiducial model is shown in blue.}
    \label{fig:comp_indipv}
\end{figure}

$N_\mathrm{pass}$ increases from 3 to 6 as we improve the calculation of the IGM metallicity if we take all five physical quantities into account. From tests where we exclude one of the physical quantities, an increase of $N_\mathrm{pass}$ is shown except for when we exclude the mean stellar $\feh$. This improvement of the model can also be observed in Fig.~\ref{fig:hist_fid}. 
However, we also find some of the \ctp trees actually have lower p-values from the improved model than from the fiducial one (Fig.~\ref{fig:comp_indipv}). 
Despite the fact that the overall range of $\langle \feh \rangle$ is now more similar between the observation and the improved model, the improved model still does not reproduce the cumulative distribution completely. This could explain why not all p-values from the 30 \ctp trees improve. Moreover, the new calculation of IGM metallicity also affects the metallicity distribution function of the MW, which is one of the observables that were used to calibrate the \asloth model. We plan to further improve the metal mixing model in a future investigation. 

To confirm that the above findings are independent of the p-value threshold, we test 4 additional cases where the p-value threshold = [0.003, 0.005, 0.03, 0.05]. We draw the same conclusion that mean stellar [Fe/H] is the property that our fiducial A-SLOTH model fails to reproduce. The improvement of the physical model is present regardless of the choice of p-value threshold. The number of \ctp trees that pass the threshold from these 4 tests are shown in Tables.~\ref{tab:excl_npass0003}-\ref{tab:excl_npass005}.

\subsection{p-value vs. properties of the MW-like galaxies}
\begin{figure*}
    \centering
    \includegraphics[width=0.9\textwidth]{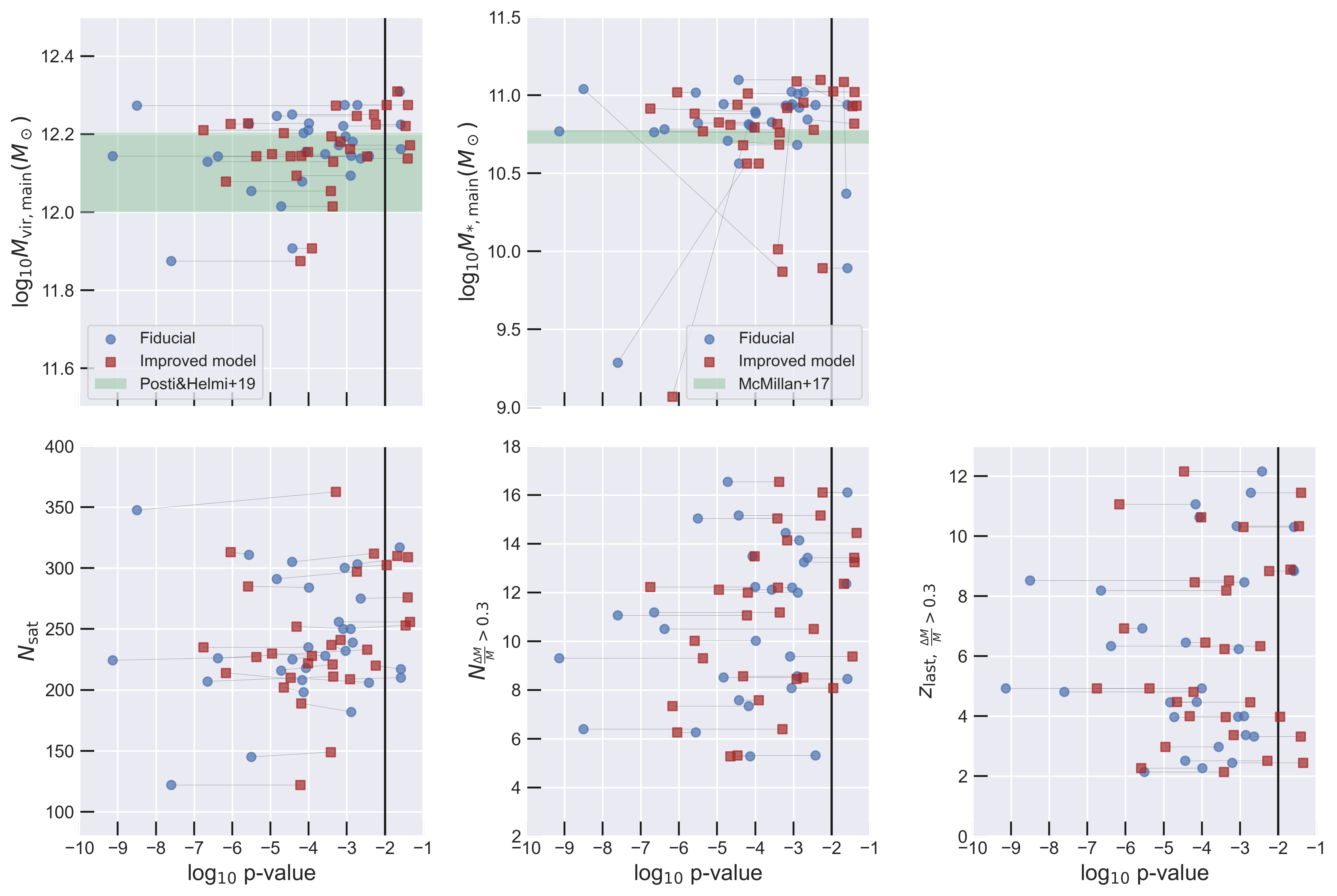}
    \caption[p-value vs. properties of MW-like galaxies from 30 \ctp trees]{ The p-values of individual \ctp trees vs. the properties of the main halo: the virial mass at $z=0$, the stellar mass at $z=0$, number of satellites (after applying the selection function), number of major halo growths ($\Delta M/M >30\%$), and the redshift of most recent major halo growth. Data from the fiducial model and the improved model are shown in blue and brown, respectively. The pairs of data points from the same trees are connected with thin black lines. The data is shifted slightly in the y-axis such that the pairs can be more easily identified. The observed MW stellar mass \citep{McMillan:2017aa} and virial mass \citep{Posti:2019aa} are added with upper and lower limits in green bands. } 
    \label{fig:pv_gal}
\end{figure*}

As discussed in Sec.~\ref{sec:ml_fidres}, the p-values from our 30 \ctp trees span a wide range. This is expected because our analysis involves information from the satellites, whereas the selection criteria for MW-like merger trees are only related with the main halo. In this section we further analyse the results regarding other properties of the dark matter merger trees to see if we can find consensus among the trees that have p-values larger than 0.01, below which we reject the null hypothesis that the tree is similar to the MW system.
We show the p-values of individual \ctp trees from both the fiducial model and the improved one vs. some properties of the main galaxies (the MWs) in the trees: the virial mass of the halo at $z=0$, the stellar mass at $z=0$, number of satellites after applying the selection function, number of major halo growth ($\Delta M/M >30\%$), and the redshift of most recent major halo growth in Fig.~\ref{fig:pv_gal}. 
Compared to the trees that have p-values below the threshold, the virial mass of the main haloes are in a narrower range in trees that pass the threshold. We also find that these virial masses are at the upper limit of or higher than the observed MW virial mass \citep{Posti:2019aa}. 
For the other four properties, we find no distinct differences between the trees that pass the threshold and those that fail.

\subsection{Caveats}
\label{sec:ml_caveat}
The stars listed in the SAGA database do not come from one survey or one group and may not be complete. As discussed in \citet{Suda:2017aa}, some stars are detected by different groups and may have different abundances in the literature. Values with the highest ``priority parameter" are the fiducial values, which are the ones we take in our analysis. The resolving power, the publication year, whether ionisation state or molecules are used, the uncertainty, and the upper/lower limit are taken into account to determine the priority parameter.
We also note that there are only 29 MW satellites listed in the SAGA database, while there are more than 50 MW satellites found \citep{Munoz:2018aa}. We aim to extend the study to individual stars in the dwarf satellites, therefore, the SAGA database is used. After applying the selection function, there are 25 observed MW satellites and an average of $> 200$ MW satellites from the \ctp trees. This leads to the long-standing issue with the $\Lambda$CDM simulations, the ``missing satellite problem" \citep{Kauffmann:1993aa,Moore:1999aa,Klypin:1999aa}, which is beyond the scope of this analysis and the physical models in \asloth.
\section{Conclusion}
\label{sec:ml_conclu}
In this work, we introduce a new analysis method that helps us analyse the results from the fiducial model of our semi-analytic code \asloth. Unlike other earlier studies, we are able to calibrate the model with multiple observables in one go. The observed and simulated satellites are clustered in a 5-dimensional space using the unsupervised Agglomerative hierarchical clustering algorithm. We obtain a p-value based on the clustering result from the Fisher-Freeman-Halton exact test, which tells us whether the observed and simulated satellites come from the same underlying distribution.

We first test our method with two two-dimensional Gaussian distributions to compare results from different unsupervised clustering algorithms (KMeans, Agglomerative, Spectral, and BIRCH) and study the dependence on the number of clusters. 
KMeans is the most straightforward algorithm to use. However, one needs to presume the number of clusters and the pursuit of even sizes of clusters can lead to non-intuitive result. In contrary, the algorithm of Agglomerative hierarchical clustering does not depend on the number of clusters. It builds up a dendrogram and depending on the number of clusters requested, it returns the labelled data points based on the dendrogram.
When the two Gaussian distributions are separated by $1\,\sigma$, we find sufficiently low p-values that allow us to reject the null hypothesis, which assumes that the two distributions come from the same underlying distribution.
We observe a decrease in p-values when the number of clusters ($N_\mathrm{cl}$) increases. At $N_\mathrm{cl} \geq 5$, we start to observe converged p-values. Therefore, we adopt Agglomerative with 5 clusters as the fiducial values.

We then apply this method to the \asloth simulated MW satellites and the observed ones. There are 5 physical quantities used in the analysis: the stellar mass $M_*$, the heliocentric distance $D_\odot$, the virial mass $M_\mathrm{vir}$, mean stellar metallicity $\langle \feh \rangle$, and the scatter among the stellar metallicity $\sigma_\feh$. 
The simulation is run with the fiducial model in the semi-analytic code \asloth and we use 30 \ctp trees \citep{Griffen:2016aa}. 
Due to the limitation of spatial information within the halo, we sample the solar position and run the analysis 100 times. The geometric mean of the 100 p-values is taken as the final p-value, which is $10^{-3.50 \pm 1.42}$ from our fiducial \asloth model. This tells us that the simulated MW satellites from our fiducial model do not come from the same underlying distribution of the observed ones, i.e., the physical model in \asloth is not good enough. 

We further analyse the simulated MW satellites and find that although the fiducial model in \asloth is able to reproduce the stellar mass-to-halo mass relation at $z=0$ well, it is not able to reproduce the chemical properties of the observed MW satellites. 
We define $N_\mathrm{pass}$ as the number of \ctp trees that have p-values > 0.01, below which we reject the null hypothesis that the tree is similar to the MW system. Tests where we exclude one of the five quantities at a time indicate that the mean stellar $\feh$ is the main quantity that \asloth fails to reproduce (Table~\ref{tab:excl_npass}). In Fig.~\ref{fig:hist_fid}, we find that most of the \asloth simulated satellites have stellar $\feh$ almost 1 dex lower than the observed ones.
This can be partly explained by the simplified assumption of homogeneous and instantaneous mixing when we determine the matallicity of the re-accreted IGM. 
We implement an ad-hoc model that assumes the IGM metallicity is higher in the proximity to the halo and the metallicity of the accreted gas is 8 times the mean IGM value.
The goal of this paper is to introduce the new analysis method and we will further improve the chemical model in \asloth in future works.

To understand why some of the \ctp trees pass the $p = 0.01$ threshold while the others don't, we further compare the p-values from individual trees with the properties of their main haloes: the virial mass at $z=0$, the stellar mass at $z=0$, the number of satellites after applying the selection function, number of major halo growths ($\Delta M/M > 30\%$), and the redshift of most recent major halo growth. 
Among the trees that pass the threshold, the virial masses of their main halo are in a narrow range [$10^{12.1} \msun - 10^{12.3} \msun$]. 
For the other properties, there are no distinct differences between the trees that pass the threshold and those that fail.

This new method of comparing observational and simulated data in high-dimensional space can distinguish a good model from a bad one easily. It has no limitation on the data size or how the data distribution looks like. 
We aim to further improve the physical model in \asloth and continue using this method.
More importantly, we plan to consider stellar information such as [C/Fe], [Ba/Fe], [Eu/Fe], etc, to study the elemental abundances of individual stars of the MW satellites in our future works.

\section*{Acknowledgements}
We thank Kohei Hayashi, Victor Ksoll, Mattis Magg, Takuma Suda, and  Yi-Shan Wu for useful discussions, and we thank the referee for highly constructive feedback. 
We gratefully acknowledge the HPC resources and data storage service SDS@hd supported by the Ministry of Science, Research and the Arts Baden-Württemberg (MWK) and the German Research Foundation (DFG) through grant INST 35/1314-1 FUGG and INST 35/1503-1 FUGG.
TH acknowledges funding from JSPS KAKENHI Grant Numbers 19K23437 and 20K14464. LHC, RSK, and SCOG are thankful for financial support from DFG via the Collaborative Research Center ’The Milky Way System’ (SFB 881, Funding-ID 138713538, subprojects A1, B1, B2, B8). RSK furthermore acknowledges support from the Heidelberg Cluster of Excellence `STRUCTURES' (EXC 2181 - 390900948), and from the European Research Council in the ERC Synergy Grant `ECOGAL' (project ID 855130). Machine learning efforts in the group are also supported by the German Space Agency (DLR) and the Federal Ministry for Economic Affairs and Climate Action (BMWK) in project `MAINN' (grant number 50OO2206). 

\section*{Software}
\asloth \citep{Magg:2022ac}, matplotlib \citep{Hunter:2007aa}, numpy \citep{Harris:2020aa}, pandas \citep{McKinney:2010aa,Reback:2022aa}, python \citep{python09}, Scikit-learn \citep{scikit-learn}, scipy \citep{Virtanen:2020aa}, stats in R \citep{Rstats:2013aa,R:2021aa}.

\section*{DATA AVAILABILITY}
The observational data used in this work can be found in the references listed in the main text. The simulated data and analysis results underlying this article will be shared on reasonable request to the corresponding author. \asloth is publicly available at \href{https://gitlab.com/thartwig/asloth}{https://gitlab.com/thartwig/asloth} and the analysis script used in this work can be found at \href{https://gitlab.com/thartwig/asloth/-/tree/main/scripts/p-value}{https://gitlab.com/thartwig/asloth/-/tree/main/scripts/p-value}.

\bibliographystyle{mnras}
\bibliography{Reference.bib}

\begin{thebibliography}{}
\makeatletter
\relax
\def\mn@urlcharsother{\let\do\@makeother \do\$\do\&\do\#\do\^\do\_\do\%\do\~}
\def\mn@doi{\begingroup\mn@urlcharsother \@ifnextchar [ {\mn@doi@}
  {\mn@doi@[]}}
\def\mn@doi@[#1]#2{\def\@tempa{#1}\ifx\@tempa\@empty \href
  {http://dx.doi.org/#2} {doi:#2}\else \href {http://dx.doi.org/#2} {#1}\fi
  \endgroup}
\def\mn@eprint#1#2{\mn@eprint@#1:#2::\@nil}
\def\mn@eprint@arXiv#1{\href {http://arxiv.org/abs/#1} {{\tt arXiv:#1}}}
\def\mn@eprint@dblp#1{\href {http://dblp.uni-trier.de/rec/bibtex/#1.xml}
  {dblp:#1}}
\def\mn@eprint@#1:#2:#3:#4\@nil{\def\@tempa {#1}\def\@tempb {#2}\def\@tempc
  {#3}\ifx \@tempc \@empty \let \@tempc \@tempb \let \@tempb \@tempa \fi \ifx
  \@tempb \@empty \def\@tempb {arXiv}\fi \@ifundefined
  {mn@eprint@\@tempb}{\@tempb:\@tempc}{\expandafter \expandafter \csname
  mn@eprint@\@tempb\endcsname \expandafter{\@tempc}}}

\bibitem[\protect\citeauthoryear{{Battaglia} \& {Nipoti}}{{Battaglia} \&
  {Nipoti}}{2022}]{Battaglia:2022ab}
{Battaglia} G.,  {Nipoti} C.,  2022, \mn@doi [Nature Astronomy]
  {10.1038/s41550-022-01638-7}, \href
  {https://ui.adsabs.harvard.edu/abs/2022NatAs.tmp..103B} {6, 659}

\bibitem[\protect\citeauthoryear{{Battaglia}, {Taibi}, {Thomas}  \&
  {Fritz}}{{Battaglia} et~al.}{2022}]{Battaglia:2022aa}
{Battaglia} G.,  {Taibi} S.,  {Thomas} G.~F.,   {Fritz} T.~K.,  2022, \mn@doi
  [\aap] {10.1051/0004-6361/202141528}, \href
  {https://ui.adsabs.harvard.edu/abs/2022A&A...657A..54B} {657, A54}

\bibitem[\protect\citeauthoryear{{Belokurov} et~al.,}{{Belokurov}
  et~al.}{2010}]{Belokurov:2010aa}
{Belokurov} V.,  et~al., 2010, \mn@doi [\apjl] {10.1088/2041-8205/712/1/L103},
  \href {https://ui.adsabs.harvard.edu/abs/2010ApJ...712L.103B} {712, L103}

\bibitem[\protect\citeauthoryear{{Chen}, {Magg}, {Hartwig}, {Glover}, {Ji}  \&
  {Klessen}}{{Chen} et~al.}{2022}]{lhc:2022aa}
{Chen} L.-H.,  {Magg} M.,  {Hartwig} T.,  {Glover} S. C.~O.,  {Ji} A.~P.,
  {Klessen} R.~S.,  2022, \mn@doi [\mnras] {10.1093/mnras/stac933}, \href
  {https://ui.adsabs.harvard.edu/abs/2022MNRAS.513..934C} {513, 934}

\bibitem[\protect\citeauthoryear{{Chiti}, {Frebel}, {Ji}, {Jerjen}, {Kim}  \&
  {Norris}}{{Chiti} et~al.}{2018}]{Chiti:2018aa}
{Chiti} A.,  {Frebel} A.,  {Ji} A.~P.,  {Jerjen} H.,  {Kim} D.,   {Norris}
  J.~E.,  2018, \mn@doi [\apj] {10.3847/1538-4357/aab4fc}, \href
  {https://ui.adsabs.harvard.edu/abs/2018ApJ...857...74C} {857, 74}

\bibitem[\protect\citeauthoryear{{Chiti} et~al.,}{{Chiti}
  et~al.}{2021}]{Chiti:2021ac}
{Chiti} A.,  et~al., 2021, \mn@doi [Nature Astronomy]
  {10.1038/s41550-020-01285-w}, \href
  {https://ui.adsabs.harvard.edu/abs/2021NatAs...5..392C} {5, 392}

\bibitem[\protect\citeauthoryear{{Chiti} et~al.,}{{Chiti}
  et~al.}{2022}]{Chiti:2022aa}
{Chiti} A.,  et~al., 2022, arXiv e-prints, \href
  {https://ui.adsabs.harvard.edu/abs/2022arXiv220501740C} {p. arXiv:2205.01740}

\bibitem[\protect\citeauthoryear{{Drlica-Wagner} et~al.,}{{Drlica-Wagner}
  et~al.}{2015}]{Drlica-Wagner:2015aa}
{Drlica-Wagner} A.,  et~al., 2015, \mn@doi [\apj]
  {10.1088/0004-637X/813/2/109}, \href
  {https://ui.adsabs.harvard.edu/abs/2015ApJ...813..109D} {813, 109}

\bibitem[\protect\citeauthoryear{{Errani}, {Pe{\~n}arrubia}  \&
  {Walker}}{{Errani} et~al.}{2018}]{Errani:2018aa}
{Errani} R.,  {Pe{\~n}arrubia} J.,   {Walker} M.~G.,  2018, \mn@doi [\mnras]
  {10.1093/mnras/sty2505}, \href
  {https://ui.adsabs.harvard.edu/abs/2018MNRAS.481.5073E} {481, 5073}

\bibitem[\protect\citeauthoryear{{Fasano} \& {Franceschini}}{{Fasano} \&
  {Franceschini}}{1987}]{Fasano:1987aa}
{Fasano} G.,  {Franceschini} A.,  1987, \mn@doi [\mnras]
  {10.1093/mnras/225.1.155}, 225, 155

\bibitem[\protect\citeauthoryear{Fisher}{Fisher}{1934}]{Fisher:1934aa}
Fisher R.~A.,  1934, Statistical Methods for Research Workers, Fifth Edision.
Oliver and Boyd, Edingburgh

\bibitem[\protect\citeauthoryear{{Font} et~al.,}{{Font}
  et~al.}{2011}]{Font:2011aa}
{Font} A.~S.,  et~al., 2011, \mn@doi [\mnras]
  {10.1111/j.1365-2966.2011.19339.x}, \href
  {https://ui.adsabs.harvard.edu/abs/2011MNRAS.417.1260F} {417, 1260}

\bibitem[\protect\citeauthoryear{Freeman \& Halton}{Freeman \&
  Halton}{1951}]{Freeman:1951aa}
Freeman G.~H.,  Halton J.~H.,  1951, Biometrika, 38, 141

\bibitem[\protect\citeauthoryear{{Gallart} et~al.,}{{Gallart}
  et~al.}{2021}]{Gallart:2021aa}
{Gallart} C.,  et~al., 2021, \mn@doi [\apj] {10.3847/1538-4357/abddbe}, \href
  {https://ui.adsabs.harvard.edu/abs/2021ApJ...909..192G} {909, 192}

\bibitem[\protect\citeauthoryear{{Garcia-Dias}, {Allende Prieto}, {S{\'a}nchez
  Almeida}  \& {Ordov{\'a}s-Pascual}}{{Garcia-Dias}
  et~al.}{2018}]{Garcia-Dias:2018aa}
{Garcia-Dias} R.,  {Allende Prieto} C.,  {S{\'a}nchez Almeida} J.,
  {Ordov{\'a}s-Pascual} I.,  2018, \mn@doi [\aap]
  {10.1051/0004-6361/201732134}, \href
  {https://ui.adsabs.harvard.edu/abs/2018A&A...612A..98G} {612, A98}

\bibitem[\protect\citeauthoryear{{Griffen}, {Ji}, {Dooley}, {G{\'o}mez},
  {Vogelsberger}, {O'Shea}  \& {Frebel}}{{Griffen}
  et~al.}{2016}]{Griffen:2016aa}
{Griffen} B.~F.,  {Ji} A.~P.,  {Dooley} G.~A.,  {G{\'o}mez} F.~A.,
  {Vogelsberger} M.,  {O'Shea} B.~W.,   {Frebel} A.,  2016, \mn@doi [\apj]
  {10.3847/0004-637X/818/1/10}, \href
  {https://ui.adsabs.harvard.edu/abs/2016ApJ...818...10G} {818, 10}

\bibitem[\protect\citeauthoryear{Harris et~al.,}{Harris
  et~al.}{2020}]{Harris:2020aa}
Harris C.~R.,  et~al., 2020, \mn@doi [Nature] {10.1038/s41586-020-2649-2}, 585,
  357

\bibitem[\protect\citeauthoryear{{Hartwig}, {Bromm}, {Klessen}  \&
  {Glover}}{{Hartwig} et~al.}{2015}]{Hartwig:2015aa}
{Hartwig} T.,  {Bromm} V.,  {Klessen} R.~S.,   {Glover} S.~C.~O.,  2015,
  \mn@doi [\mnras] {10.1093/mnras/stu2740}, \href
  {http://adsabs.harvard.edu/abs/2015MNRAS.447.3892H} {447, 3892}

\bibitem[\protect\citeauthoryear{{Hartwig} et~al.,}{{Hartwig}
  et~al.}{2022}]{Hartwig:2022aa}
{Hartwig} T.,  et~al., 2022, \mn@doi [\apj] {10.3847/1538-4357/ac7150}, \href
  {https://ui.adsabs.harvard.edu/abs/2022ApJ...936...45H} {936, 45}

\bibitem[\protect\citeauthoryear{Hunter}{Hunter}{2007}]{Hunter:2007aa}
Hunter J.~D.,  2007, \mn@doi [Computing in Science \& Engineering]
  {10.1109/MCSE.2007.55}, 9, 90

\bibitem[\protect\citeauthoryear{{Jeon}, {Besla}  \& {Bromm}}{{Jeon}
  et~al.}{2017}]{Jeon:2017aa}
{Jeon} M.,  {Besla} G.,   {Bromm} V.,  2017, \mn@doi [\apj]
  {10.3847/1538-4357/aa8c80}, \href
  {http://adsabs.harvard.edu/abs/2017ApJ...848...85J} {848, 85}

\bibitem[\protect\citeauthoryear{{Ji}, {Frebel}, {Chiti}  \& {Simon}}{{Ji}
  et~al.}{2016}]{Ji:2016aa}
{Ji} A.~P.,  {Frebel} A.,  {Chiti} A.,   {Simon} J.~D.,  2016, \mn@doi [\nat]
  {10.1038/nature17425}, \href
  {https://ui.adsabs.harvard.edu/abs/2016Natur.531..610J} {531, 610}

\bibitem[\protect\citeauthoryear{{Ji} et~al.,}{{Ji} et~al.}{2021}]{Ji:2021aa}
{Ji} A.~P.,  et~al., 2021, \mn@doi [\apj] {10.3847/1538-4357/ac1869}, \href
  {https://ui.adsabs.harvard.edu/abs/2021ApJ...921...32J} {921, 32}

\bibitem[\protect\citeauthoryear{{Kang}, {Pellegrini}, {Ardizzone}, {Klessen},
  {Koethe}, {Glover}  \& {Ksoll}}{{Kang} et~al.}{2022}]{Kang:2022aa}
{Kang} D.~E.,  {Pellegrini} E.~W.,  {Ardizzone} L.,  {Klessen} R.~S.,  {Koethe}
  U.,  {Glover} S. C.~O.,   {Ksoll} V.~F.,  2022, \mn@doi [\mnras]
  {10.1093/mnras/stac222}, \href
  {https://ui.adsabs.harvard.edu/abs/2022MNRAS.512..617K} {512, 617}

\bibitem[\protect\citeauthoryear{{Kauffmann}, {White}  \&
  {Guiderdoni}}{{Kauffmann} et~al.}{1993}]{Kauffmann:1993aa}
{Kauffmann} G.,  {White} S.~D.~M.,   {Guiderdoni} B.,  1993, \mn@doi [\mnras]
  {10.1093/mnras/264.1.201}, \href
  {https://ui.adsabs.harvard.edu/abs/1993MNRAS.264..201K} {264, 201}

\bibitem[\protect\citeauthoryear{{Kirby}, {Cohen}, {Simon}, {Guhathakurta},
  {Thygesen}  \& {Duggan}}{{Kirby} et~al.}{2017}]{Kirby:2017aa}
{Kirby} E.~N.,  {Cohen} J.~G.,  {Simon} J.~D.,  {Guhathakurta} P.,  {Thygesen}
  A.~O.,   {Duggan} G.~E.,  2017, \mn@doi [\apj] {10.3847/1538-4357/aa6570},
  \href {https://ui.adsabs.harvard.edu/abs/2017ApJ...838...83K} {838, 83}

\bibitem[\protect\citeauthoryear{{Klypin}, {Gottl{\"o}ber}, {Kravtsov}  \&
  {Khokhlov}}{{Klypin} et~al.}{1999}]{Klypin:1999aa}
{Klypin} A.,  {Gottl{\"o}ber} S.,  {Kravtsov} A.~V.,   {Khokhlov} A.~M.,  1999,
  \mn@doi [\apj] {10.1086/307122}, \href
  {https://ui.adsabs.harvard.edu/abs/1999ApJ...516..530K} {516, 530}

\bibitem[\protect\citeauthoryear{{Kolmogorov}}{{Kolmogorov}}{1933}]{Kolmogorov:1933aa}
{Kolmogorov} A.~N.,  1933, Giornale dell'Istituto Italiano degli Attuari, 4, 83

\bibitem[\protect\citeauthoryear{{Koposov}, {Yoo}, {Rix}, {Weinberg},
  {Macci{\`o}}  \& {Escud{\'e}}}{{Koposov} et~al.}{2009}]{Koposov:2009aa}
{Koposov} S.~E.,  {Yoo} J.,  {Rix} H.-W.,  {Weinberg} D.~H.,  {Macci{\`o}}
  A.~V.,   {Escud{\'e}} J.~M.,  2009, \mn@doi [\apj]
  {10.1088/0004-637X/696/2/2179}, \href
  {https://ui.adsabs.harvard.edu/abs/2009ApJ...696.2179K} {696, 2179}

\bibitem[\protect\citeauthoryear{{Koposov}, {Belokurov}, {Torrealba}  \&
  {Evans}}{{Koposov} et~al.}{2015}]{Koposov:2015aa}
{Koposov} S.~E.,  {Belokurov} V.,  {Torrealba} G.,   {Evans} N.~W.,  2015,
  \mn@doi [\apj] {10.1088/0004-637X/805/2/130}, \href
  {https://ui.adsabs.harvard.edu/abs/2015ApJ...805..130K} {805, 130}

\bibitem[\protect\citeauthoryear{{Ksoll} et~al.,}{{Ksoll}
  et~al.}{2021a}]{Ksoll:2021aa}
{Ksoll} V.~F.,  et~al., 2021a, \mn@doi [\aj] {10.3847/1538-3881/abee8b}, \href
  {https://ui.adsabs.harvard.edu/abs/2021AJ....161..256K} {161, 256}

\bibitem[\protect\citeauthoryear{{Ksoll} et~al.,}{{Ksoll}
  et~al.}{2021b}]{Ksoll:2021ab}
{Ksoll} V.~F.,  et~al., 2021b, \mn@doi [\aj] {10.3847/1538-3881/abee8c}, \href
  {https://ui.adsabs.harvard.edu/abs/2021AJ....161..257K} {161, 257}

\bibitem[\protect\citeauthoryear{{Logan} \& {Fotopoulou}}{{Logan} \&
  {Fotopoulou}}{2020}]{Logan:2020aa}
{Logan} C.~H.~A.,  {Fotopoulou} S.,  2020, \mn@doi [\aap]
  {10.1051/0004-6361/201936648}, \href
  {https://ui.adsabs.harvard.edu/abs/2020A&A...633A.154L} {633, A154}

\bibitem[\protect\citeauthoryear{{Magg}, {Hartwig}, {Glover}, {Klessen}  \&
  {Whalen}}{{Magg} et~al.}{2016}]{Magg:2016aa}
{Magg} M.,  {Hartwig} T.,  {Glover} S.~C.~O.,  {Klessen} R.~S.,   {Whalen}
  D.~J.,  2016, \mn@doi [\mnras] {10.1093/mnras/stw1882}, \href
  {http://adsabs.harvard.edu/abs/2016MNRAS.462.3591M} {462, 3591}

\bibitem[\protect\citeauthoryear{{Magg}, {Hartwig}, {Chen}  \& {Tarumi}}{{Magg}
  et~al.}{2022}]{Magg:2022ac}
{Magg} M.,  {Hartwig} T.,  {Chen} L.-H.,   {Tarumi} Y.,  2022, \mn@doi [The
  Journal of Open Source Software] {10.21105/joss.04417}, \href
  {https://ui.adsabs.harvard.edu/abs/2022JOSS....7.4417M} {7, 4417}

\bibitem[\protect\citeauthoryear{{McConnachie}}{{McConnachie}}{2012}]{McConnachie:2012aa}
{McConnachie} A.~W.,  2012, \mn@doi [\aj] {10.1088/0004-6256/144/1/4}, \href
  {https://ui.adsabs.harvard.edu/abs/2012AJ....144....4M} {144, 4}

\bibitem[\protect\citeauthoryear{{McConnachie} \& {Venn}}{{McConnachie} \&
  {Venn}}{2020}]{McConnachie:2020aa}
{McConnachie} A.~W.,  {Venn} K.~A.,  2020, \mn@doi [\aj]
  {10.3847/1538-3881/aba4ab}, \href
  {https://ui.adsabs.harvard.edu/abs/2020AJ....160..124M} {160, 124}

\bibitem[\protect\citeauthoryear{{McMillan}}{{McMillan}}{2017}]{McMillan:2017aa}
{McMillan} P.~J.,  2017, \mn@doi [\mnras] {10.1093/mnras/stw2759}, \href
  {https://ui.adsabs.harvard.edu/abs/2017MNRAS.465...76M} {465, 76}

\bibitem[\protect\citeauthoryear{{Moore}, {Ghigna}, {Governato}, {Lake},
  {Quinn}, {Stadel}  \& {Tozzi}}{{Moore} et~al.}{1999}]{Moore:1999aa}
{Moore} B.,  {Ghigna} S.,  {Governato} F.,  {Lake} G.,  {Quinn} T.,  {Stadel}
  J.,   {Tozzi} P.,  1999, \mn@doi [\apjl] {10.1086/312287}, \href
  {https://ui.adsabs.harvard.edu/abs/1999ApJ...524L..19M} {524, L19}

\bibitem[\protect\citeauthoryear{{Mu{\~n}oz}, {Carlin}, {Frinchaboy},
  {Nidever}, {Majewski}  \& {Patterson}}{{Mu{\~n}oz}
  et~al.}{2006}]{Munoz:2006aa}
{Mu{\~n}oz} R.~R.,  {Carlin} J.~L.,  {Frinchaboy} P.~M.,  {Nidever} D.~L.,
  {Majewski} S.~R.,   {Patterson} R.~J.,  2006, \mn@doi [\apjl]
  {10.1086/508685}, \href
  {https://ui.adsabs.harvard.edu/abs/2006ApJ...650L..51M} {650, L51}

\bibitem[\protect\citeauthoryear{{Mu{\~n}oz}, {C{\^o}t{\'e}}, {Santana},
  {Geha}, {Simon}, {Oyarz{\'u}n}, {Stetson}  \& {Djorgovski}}{{Mu{\~n}oz}
  et~al.}{2018}]{Munoz:2018aa}
{Mu{\~n}oz} R.~R.,  {C{\^o}t{\'e}} P.,  {Santana} F.~A.,  {Geha} M.,  {Simon}
  J.~D.,  {Oyarz{\'u}n} G.~A.,  {Stetson} P.~B.,   {Djorgovski} S.~G.,  2018,
  \mn@doi [\apj] {10.3847/1538-4357/aac16b}, \href
  {https://ui.adsabs.harvard.edu/abs/2018ApJ...860...66M} {860, 66}

\bibitem[\protect\citeauthoryear{{Peacock}}{{Peacock}}{1983}]{Peacock:1983aa}
{Peacock} J.~A.,  1983, \mn@doi [\mnras] {10.1093/mnras/202.3.615}, \href
  {https://ui.adsabs.harvard.edu/abs/1983MNRAS.202..615P} {202, 615}

\bibitem[\protect\citeauthoryear{{Pearson}}{{Pearson}}{1916}]{Peasron:1916aa}
{Pearson} K.,  1916, \mn@doi [Philosophical Transactions of the Royal Society
  of London Series A] {10.1098/rsta.1916.0009}, \href
  {https://ui.adsabs.harvard.edu/abs/1916RSPTA.216..429P} {216, 429}

\bibitem[\protect\citeauthoryear{Pedregosa et~al.,}{Pedregosa
  et~al.}{2011}]{scikit-learn}
Pedregosa F.,  et~al., 2011, Journal of Machine Learning Research, 12, 2825

\bibitem[\protect\citeauthoryear{{Posti} \& {Helmi}}{{Posti} \&
  {Helmi}}{2019}]{Posti:2019aa}
{Posti} L.,  {Helmi} A.,  2019, \mn@doi [\aap] {10.1051/0004-6361/201833355},
  \href {https://ui.adsabs.harvard.edu/abs/2019A&A...621A..56P} {621, A56}

\bibitem[\protect\citeauthoryear{{R Core Team}}{{R Core
  Team}}{2013}]{Rstats:2013aa}
{R Core Team} 2013, R: A Language and Environment for Statistical Computing.
R Foundation for Statistical Computing, Vienna, Austria, \url
  {http://www.R-project.org/}

\bibitem[\protect\citeauthoryear{{R Core Team}}{{R Core Team}}{2021}]{R:2021aa}
{R Core Team} 2021, R: A Language and Environment for Statistical Computing.
R Foundation for Statistical Computing, Vienna, Austria, \url
  {https://www.R-project.org/}

\bibitem[\protect\citeauthoryear{{Reback} et~al.,}{{Reback}
  et~al.}{2022}]{Reback:2022aa}
{Reback} J.,  et~al., 2022, pandas-dev/pandas: Pandas 1.4.1,
  \mn@doi{10.5281/zenodo.6053272.
}

\bibitem[\protect\citeauthoryear{{Reichert}, {Hansen}, {Hanke},
  {Sk{\'u}lad{\'o}ttir}, {Arcones}  \& {Grebel}}{{Reichert}
  et~al.}{2020}]{Reichert:2020aa}
{Reichert} M.,  {Hansen} C.~J.,  {Hanke} M.,  {Sk{\'u}lad{\'o}ttir} {\'A}.,
  {Arcones} A.,   {Grebel} E.~K.,  2020, \mn@doi [\aap]
  {10.1051/0004-6361/201936930}, \href
  {https://ui.adsabs.harvard.edu/abs/2020A&A...641A.127R} {641, A127}

\bibitem[\protect\citeauthoryear{{Reis}, {Rotman}, {Poznanski}, {Prochaska}  \&
  {Wolf}}{{Reis} et~al.}{2019}]{Reis:2019aa}
{Reis} I.,  {Rotman} M.,  {Poznanski} D.,  {Prochaska} J.~X.,   {Wolf} L.,
  2019, arXiv e-prints, \href
  {https://ui.adsabs.harvard.edu/abs/2019arXiv191106823R} {p. arXiv:1911.06823}

\bibitem[\protect\citeauthoryear{{Ricotti} \& {Gnedin}}{{Ricotti} \&
  {Gnedin}}{2005}]{Ricotti:2005aa}
{Ricotti} M.,  {Gnedin} N.~Y.,  2005, \mn@doi [\apj] {10.1086/431415}, \href
  {https://ui.adsabs.harvard.edu/abs/2005ApJ...629..259R} {629, 259}

\bibitem[\protect\citeauthoryear{{Rolleston}, {Venn}, {Tolstoy}  \&
  {Dufton}}{{Rolleston} et~al.}{2003}]{Rolleston:2003aa}
{Rolleston} W.~R.~J.,  {Venn} K.,  {Tolstoy} E.,   {Dufton} P.~L.,  2003,
  \mn@doi [\aap] {10.1051/0004-6361:20021653}, \href
  {https://ui.adsabs.harvard.edu/abs/2003A&A...400...21R} {400, 21}

\bibitem[\protect\citeauthoryear{{Romano}, {Calura}, {D'Ercole}  \&
  {Few}}{{Romano} et~al.}{2019}]{Romano:2019aa}
{Romano} D.,  {Calura} F.,  {D'Ercole} A.,   {Few} C.~G.,  2019, \mn@doi [\aap]
  {10.1051/0004-6361/201935328}, \href
  {https://ui.adsabs.harvard.edu/abs/2019A&A...630A.140R} {630, A140}

\bibitem[\protect\citeauthoryear{{Safarzadeh} \& {Scannapieco}}{{Safarzadeh} \&
  {Scannapieco}}{2017}]{Safarzadeh:2017aa}
{Safarzadeh} M.,  {Scannapieco} E.,  2017, \mn@doi [\mnras]
  {10.1093/mnras/stx1706}, \href
  {https://ui.adsabs.harvard.edu/abs/2017MNRAS.471.2088S} {471, 2088}

\bibitem[\protect\citeauthoryear{{Salvadori}, {Sk{\'u}lad{\'o}ttir}  \&
  {Tolstoy}}{{Salvadori} et~al.}{2015}]{Salvadori:2015aa}
{Salvadori} S.,  {Sk{\'u}lad{\'o}ttir} {\'A}.,   {Tolstoy} E.,  2015, \mn@doi
  [\mnras] {10.1093/mnras/stv1969}, \href
  {https://ui.adsabs.harvard.edu/abs/2015MNRAS.454.1320S} {454, 1320}

\bibitem[\protect\citeauthoryear{{Sanati}, {Jeanquartier}, {Revaz}  \&
  {Jablonka}}{{Sanati} et~al.}{2022}]{Sanati:2022aa}
{Sanati} M.,  {Jeanquartier} F.,  {Revaz} Y.,   {Jablonka} P.,  2022, arXiv
  e-prints, \href {https://ui.adsabs.harvard.edu/abs/2022arXiv220611351S} {p.
  arXiv:2206.11351}

\bibitem[\protect\citeauthoryear{{Simon}}{{Simon}}{2019}]{Simon:2019aa}
{Simon} J.~D.,  2019, \mn@doi [\araa] {10.1146/annurev-astro-091918-104453},
  \href {https://ui.adsabs.harvard.edu/abs/2019ARA&A..57..375S} {57, 375}

\bibitem[\protect\citeauthoryear{{Starkenburg} et~al.,}{{Starkenburg}
  et~al.}{2013}]{Starkenburg:2013aa}
{Starkenburg} E.,  et~al., 2013, \mn@doi [\mnras] {10.1093/mnras/sts367}, \href
  {https://ui.adsabs.harvard.edu/abs/2013MNRAS.429..725S} {429, 725}

\bibitem[\protect\citeauthoryear{{Suda} et~al.,}{{Suda}
  et~al.}{2008}]{Suda:2008aa}
{Suda} T.,  et~al., 2008, \mn@doi [\pasj] {10.1093/pasj/60.5.1159}, \href
  {https://ui.adsabs.harvard.edu/abs/2008PASJ...60.1159S} {60, 1159}

\bibitem[\protect\citeauthoryear{{Suda} et~al.,}{{Suda}
  et~al.}{2017}]{Suda:2017aa}
{Suda} T.,  et~al., 2017, \mn@doi [\pasj] {10.1093/pasj/psx059}, \href
  {https://ui.adsabs.harvard.edu/abs/2017PASJ...69...76S} {69, 76}

\bibitem[\protect\citeauthoryear{{Tarumi}, {Hartwig}  \& {Magg}}{{Tarumi}
  et~al.}{2020}]{Tarumi:2020ab}
{Tarumi} Y.,  {Hartwig} T.,   {Magg} M.,  2020, \mn@doi [\apj]
  {10.3847/1538-4357/ab960d}, \href
  {https://ui.adsabs.harvard.edu/abs/2020ApJ...897...58T} {897, 58}

\bibitem[\protect\citeauthoryear{{Tsujimoto}, {Nomoto}, {Yoshii}, {Hashimoto},
  {Yanagida}  \& {Thielemann}}{{Tsujimoto} et~al.}{1995}]{Tsujimoto:1995aa}
{Tsujimoto} T.,  {Nomoto} K.,  {Yoshii} Y.,  {Hashimoto} M.,  {Yanagida} S.,
  {Thielemann} F.~K.,  1995, \mn@doi [\mnras] {10.1093/mnras/277.3.945}, \href
  {https://ui.adsabs.harvard.edu/abs/1995MNRAS.277..945T} {277, 945}

\bibitem[\protect\citeauthoryear{Van~Rossum \& Drake}{Van~Rossum \&
  Drake}{2009}]{python09}
Van~Rossum G.,  Drake F.~L.,  2009, Python 3 Reference Manual.
CreateSpace, Scotts Valley, CA

\bibitem[\protect\citeauthoryear{{Van der Swaelmen}, {Hill}, {Primas}  \&
  {Cole}}{{Van der Swaelmen} et~al.}{2013}]{VanderSwaelmen:2013aa}
{Van der Swaelmen} M.,  {Hill} V.,  {Primas} F.,   {Cole} A.~A.,  2013, \mn@doi
  [\aap] {10.1051/0004-6361/201321109}, \href
  {https://ui.adsabs.harvard.edu/abs/2013A&A...560A..44V} {560, A44}

\bibitem[\protect\citeauthoryear{Virtanen et~al.,}{Virtanen
  et~al.}{2020}]{Virtanen:2020aa}
Virtanen P.,  et~al., 2020, \mn@doi [Nature Methods]
  {10.1038/s41592-019-0686-2}, \href {https://rdcu.be/b08Wh} {17, 261}

\bibitem[\protect\citeauthoryear{{Walker}, {Mateo}, {Olszewski}, {Gnedin},
  {Wang}, {Sen}  \& {Woodroofe}}{{Walker} et~al.}{2007}]{Walker:2007aa}
{Walker} M.~G.,  {Mateo} M.,  {Olszewski} E.~W.,  {Gnedin} O.~Y.,  {Wang} X.,
  {Sen} B.,   {Woodroofe} M.,  2007, \mn@doi [\apjl] {10.1086/521998}, \href
  {https://ui.adsabs.harvard.edu/abs/2007ApJ...667L..53W} {667, L53}

\bibitem[\protect\citeauthoryear{{Wang} et~al.,}{{Wang}
  et~al.}{2021}]{Wang:2021aa}
{Wang} W.,  et~al., 2021, \mn@doi [\mnras] {10.1093/mnras/staa3495}, \href
  {https://ui.adsabs.harvard.edu/abs/2021MNRAS.500.3776W} {500, 3776}

\bibitem[\protect\citeauthoryear{{Wang}, {Shi}, {Yan}, {Xia}, {Zhao}, {Li}  \&
  {Li}}{{Wang} et~al.}{2022}]{Wang:2022aa}
{Wang} G.-J.,  {Shi} H.-L.,  {Yan} Y.-P.,  {Xia} J.-Q.,  {Zhao} Y.-Y.,  {Li}
  S.-Y.,   {Li} J.-F.,  2022, \mn@doi [\apjs] {10.3847/1538-4365/ac5f4a}, \href
  {https://ui.adsabs.harvard.edu/abs/2022ApJS..260...13W} {260, 13}

\bibitem[\protect\citeauthoryear{{Weisz}, {Dolphin}, {Skillman}, {Holtzman},
  {Gilbert}, {Dalcanton}  \& {Williams}}{{Weisz} et~al.}{2014}]{Weisz:2014aa}
{Weisz} D.~R.,  {Dolphin} A.~E.,  {Skillman} E.~D.,  {Holtzman} J.,  {Gilbert}
  K.~M.,  {Dalcanton} J.~J.,   {Williams} B.~F.,  2014, \mn@doi [\apj]
  {10.1088/0004-637X/789/2/148}, \href
  {https://ui.adsabs.harvard.edu/abs/2014ApJ...789..148W} {789, 148}

\bibitem[\protect\citeauthoryear{{W}es {M}c{K}inney}{{W}es
  {M}c{K}inney}{2010}]{McKinney:2010aa}
{W}es {M}c{K}inney 2010, in {S}t\'efan van~der {W}alt {J}arrod {M}illman eds,
  {P}roceedings of the 9th {P}ython in {S}cience {C}onference. pp 56 -- 61,
  \mn@doi{10.25080/Majora-92bf1922-00a}

\bibitem[\protect\citeauthoryear{{Wheeler} et~al.,}{{Wheeler}
  et~al.}{2019}]{Wheeler:2019aa}
{Wheeler} C.,  et~al., 2019, \mn@doi [\mnras] {10.1093/mnras/stz2887}, \href
  {https://ui.adsabs.harvard.edu/abs/2019MNRAS.490.4447W} {490, 4447}

\bibitem[\protect\citeauthoryear{{Yoon}, {Whitten}, {Beers}, {Lee}, {Masseron}
  \& {Placco}}{{Yoon} et~al.}{2020}]{Yoon:2020aa}
{Yoon} J.,  {Whitten} D.~D.,  {Beers} T.~C.,  {Lee} Y.~S.,  {Masseron} T.,
  {Placco} V.~M.,  2020, \mn@doi [\apj] {10.3847/1538-4357/ab7daf}, \href
  {https://ui.adsabs.harvard.edu/abs/2020ApJ...894....7Y} {894, 7}

\makeatother
\end{thebibliography}

\appendix
\section{Number of trees that pass the p-value threshold vs. different thresholds}
The p-value threshold of 0.01 that we use in the analysis is a conventional choice (Sec.~\ref{sec:ml_fidimp}). To examine the dependence of our results on this threshold, we test additionally 4 different values (0.003, 0.005, 0.03, and 0.05). We show how many \ctp trees pass the threshold from these tests in Tables~\ref{tab:excl_npass0003}-\ref{tab:excl_npass005}. We find a consensus among the tests that mean stellar $\feh$ is the property which the fiducial \asloth  model fails to reproduce. Our modified model produces improved results in all tests, independent of the p-value threshold.
\begin{table}
    \centering
    \begin{tabular}{l|l|l}
        Excluded quantity & $N_\mathrm{pass, fid}$ & $N_\mathrm{pass, imp}$ \\
        \hline
        $M_*$ & 1 & 18 \\
        $D_\odot$ & 1 & 8 \\
        $\langle \feh \rangle$ & 29 & 27 \\
        $\sigma_{\feh}$ & 0 & 7 \\
        $M_\mathrm{vir}$ & 0 & 8 \\
        \hline
        All 5 quantities considered & 4 & 9 \\
    \end{tabular}
    \caption{ Similar to Table~\ref{tab:excl_npass} but with p-value threshold = 0.003. }
    \label{tab:excl_npass0003}
\end{table}

\begin{table}
    \centering
    \begin{tabular}{l|l|l}
        Excluded quantity & $N_\mathrm{pass, fid}$ & $N_\mathrm{pass, imp}$ \\
        \hline
        $M_*$ & 0 & 13 \\
        $D_\odot$ & 0 & 6 \\
        $\langle \feh \rangle$ & 27 & 27 \\
        $\sigma_{\feh}$ & 0 & 6 \\
        $M_\mathrm{vir}$ & 0 & 5 \\
        \hline
        All 5 quantities considered & 3 & 8 \\
    \end{tabular}
    \caption{ Similar to Table~\ref{tab:excl_npass} but with p-value threshold = 0.005. }
    \label{tab:excl_npass0005}
\end{table}

\begin{table}
    \centering
    \begin{tabular}{l|l|l}
        Excluded quantity & $N_\mathrm{pass, fid}$ & $N_\mathrm{pass, imp}$ \\
        \hline
        $M_*$ & 0 & 5 \\
        $D_\odot$ & 0 & 3 \\
        $\langle \feh \rangle$ & 23 & 17 \\
        $\sigma_{\feh}$ & 0 & 2 \\
        $M_\mathrm{vir}$ & 0 & 2 \\
        \hline
        All 5 quantities considered & 0 & 4 \\
    \end{tabular}
    \caption{ Similar to Table~\ref{tab:excl_npass} but with p-value threshold = 0.03. }
    \label{tab:excl_npass003}
\end{table}

\begin{table}
    \centering
    \begin{tabular}{l|l|l}
        Excluded quantity & $N_\mathrm{pass, fid}$ & $N_\mathrm{pass, imp}$ \\
        \hline
        $M_*$ & 0 & 3 \\
        $D_\odot$ & 0 & 3 \\
        $\langle \feh \rangle$ & 17 & 16 \\
        $\sigma_{\feh}$ & 0 & 2 \\
        $M_\mathrm{vir}$ & 0 & 2 \\
        \hline
        All 5 quantities considered & 0 & 0 \\
    \end{tabular}
    \caption{ Similar to Table~\ref{tab:excl_npass} but with p-value threshold = 0.05. }
    \label{tab:excl_npass005}
\end{table}

\bsp	
\label{lastpage}
\end{document}